\documentclass[manuscript]{acmart}
\usepackage{graphicx}
\usepackage{array}
\usepackage{adjustbox} 
\usepackage{multicol}
\usepackage{subcaption}

\AtBeginDocument{%
  }

\setcopyright{acmlicensed}
\copyrightyear{2018}
\acmYear{2025}
\acmDOI{XXXXXXX.XXXXXXX}

\begin{document}

\title{System X: A Mobile Voice-Based AI System for EMR Generation and Clinical Decision Support in Low-Resource Maternal Healthcare}

\author{Maryam Mustafa}
\email{maryam_mustafa@lums.edu.pk}
\affiliation{%
  \institution{Lahore University of Management Sciences}
  \city{Lahore}
  \country{Pakistan}
}

\author{Umme Ammara}
\email{22100100@lums.edu.pk}
\affiliation{%
  \institution{Lahore University of Management Sciences}
  \city{Lahore}
  \country{Pakistan}
}

\author{Amna Shahnawaz}
\email{amna.shahnawaz@lums.edu.pk}
\affiliation{%
  \institution{Lahore University of Management Sciences}
  \city{Lahore}
  \country{Pakistan}
}

\author{Moaiz Abrar}
\email{23100151@lums.edu.pk}
\affiliation{%
  \institution{Lahore University of Management Sciences}
  \city{Lahore}
  \country{Pakistan}
}

\author{Bakhtawar Ahtisham}
\email{24100301@lums.edu.pk}
\affiliation{%
  \institution{Lahore University of Management Sciences}
  \city{Lahore}
  \country{Pakistan}
}

\author{Fozia Umber Qurashi}
\email{fozia.umber@sihs.org.pk}
\affiliation{%
  \institution{Obstetrics \& Gynecology, Shalamar Hospital}
  \city{Lahore}
  \country{Pakistan}
}

\author{Mostafa Shahin}
\email{m.shahin@unsw.edu.au}
\affiliation{%
  \institution{University of New South Wales}
  \city{Sydney}
  \country{Australia}
}

\author{Beena Ahmed}
\email{beena.ahmed@unsw.edu.au}
\affiliation{%
  \institution{University of New South Wales}
  \city{Sydney}
  \country{Australia}
}

\renewcommand{\shortauthors}{Trovato et al.}

\begin{abstract}
We present the design, implementation, and in-situ deployment of a smartphone-based voice-enabled AI system for generating electronic medical records (EMRs) and clinical risk alerts in maternal healthcare settings. Targeted at low-resource environments such as  Pakistan, the system integrates a fine-tuned, multilingual automatic speech recognition (ASR) model and a prompt-engineered large language model (LLM) to enable healthcare workers to engage naturally in Urdu, their native language, regardless of literacy or technical background. Through speech-based input and localized understanding, the system generates structured EMRs and flags critical maternal health risks.  Over a seven-month deployment in a not-for-profit hospital, the system supported the creation of over 500 EMRs and flagged over 300 potential clinical risks. We evaluate the system’s performance across speech recognition accuracy, EMR field-level correctness, and clinical relevance of AI-generated red flags. Our results demonstrate that speech based AI interfaces, can be effectively adapted to real-world healthcare settings, especially in low-resource settings,  when combined with structured input design, contextual medical dictionaries, and clinician-in-the-loop feedback loops.  We discuss generalizable design principles for deploying voice-based mobile healthcare AI support systems in linguistically and infrastructurally constrained settings. 

\end{abstract}


\received{20 February 2007}
\received[revised]{12 March 2009}
\received[accepted]{5 June 2009}

\maketitle

\section{Introduction}
This paper presents the design, system architecture, and deployment of a smartphone-based AI assistant developed to support maternal healthcare delivery by frontline health workers in Pakistan. The assistant leverages a large language model (LLM) to generate electronic medical records (EMRs) from spoken input in the local language. By generating medical records and identifying potential health risks and flagging them in real time, the system enhances patient care while optimizing for health care workers time.

Pakistan faces one of the highest maternal mortality ratios globally, with 154 deaths per 100,000 live births, compared to an average of 12 deaths per 100,000 in developed countries~\cite{hanif2021maternal}. Several factors contribute to this high mortality rate, including the inadequate recognition of critical symptoms, incomplete or missing health records, delayed access to emergency obstetric care, and insufficiently trained healthcare workers~\cite{mumtaz2014improving}. The absence of comprehensive medical records for pregnant women in Pakistan poses significant challenges for healthcare providers, making it difficult to deliver accurate diagnoses, conduct proper follow-ups, or provide care that is contextualized based on both the patient's clinical history and socio-economic conditions~\cite{varley2023hospital}. While EMRs are widely adopted in developed countries~\cite{onc2019electronic}, most public and not-for-profit healthcare facilities in Pakistan still rely on paper-based records, with only a few major hospitals having transitioned to digital forms~\cite{malik2021needofhour}. In contrast, structured EMRs have been shown to reduce clinical errors and improve treatment accuracy in digitally mature health systems~\cite{chaudhry2006systematic, onc2019electronic}.  

Beyond cost considerations of computers/desktop devices etc. needed to run existing EMR platforms, a significant challenge in implementing EMR systems within the clinical workforce of developing countries is the issue of digital literacy~\cite{sood2004implementing}. The usability of EMRs is further compromised by difficulties in navigating interfaces and managing records through conventional keyboard and mouse interactions for healthcare workers~\cite{Ajami2013}. These usability barriers are compounded by unreliable infrastructure, high costs, and a lack of technical training~\cite{KumahCrystal2018}.

To address these challenges, in collaboration with one of the largest not-for-profit hospitals in Pakistan, we developed a mobile phone based, conversational AI assistant for maternal health workers~(Fig.\ref{fig:system-design}), to make it easier to collect digital health records, without the barriers of literacy, cost, time, training and access to equipment. The health records collect medical data, as well as data on social determinants of health factors affecting pregnancy. Our system also supports low-skilled health care workers by raising red flags based on the collected data to inform better decisions and improve patient outcomes.  The system comprises: (1) an intuitive user interface (UI) featuring structured, maternal health-specific prompts, allowing healthcare workers to provide data and record audio responses efficiently; (2) a speech recognition module utilizing a multilingual Automatic Speech Recognition (ASR) system, Whisper, to convert collected speech into the text; and (3) a text generation module prompting an existing LLM, GPT-4, with the transcribed text to populate EMRs, ask follow up questions and provide diagnostic support. The application has undergone extensive testing in the Gynecology Outpatient Department at Hospital Y, where it was deployed from October 2023 till April 2024, resulting in the creation of 502 unique patient records. Our contribution lies in how we re-engineer and contextualize existing AI models for a setting in which full-featured EMR suites remain unrealistic. Specifically, we present the country's first voice-enabled smartphone-based AI assistant for maternal care, deployed in the wild and on scale.

In this paper we address three primary research questions on the use of LLMs for improving maternal health outcomes:
\begin{itemize}
    \item \textbf{RQ~1:}  How can LLMs be effectively integrated into maternal healthcare systems in low-resource settings to generate accurate and context-sensitive EMRs from speech in local languages?

\item \textbf{RQ~2: }What are the specific challenges and opportunities associated with deploying a voice-enabled AI assistant for maternal healthcare workers (MHCWs) in Pakistan, and how effective are LLMs in generating accurate and contextually relevant EMRs from speech in low-resource healthcare settings?

\item \textbf{RQ~3:} How can LLMs enhance early detection and diagnosis of critical maternal health issues through real-time decision support for low-skilled maternal health care workers, and what potential impact can this have on improving maternal health outcomes in under-resourced areas?
\end{itemize}

To design, develop, and evaluate the system, we followed a multi-phase methodology. First, we conducted a formative study involving workflow observations and interviews with clinicians to understand existing maternal health documentation practices~(Section \ref{subsec:qual}). Next, we collected real-world Urdu speech data to fine-tune an ASR model and iteratively developed a voice-based interface integrated with a LLM for EMR generation and clinical decision support~(Section \ref{sec:system2}). We then deployed the system in a live hospital setting over seven months, capturing usage data, system logs, and user feedback~(Section \ref{sec:deployment}). Finally, we conducted a comprehensive evaluation using quantitative metrics, usability testing, and clinician interviews to assess system accuracy, utility, and impact in practice~(Section \ref{sec:evaluation}).

Our work highlights the design guidelines, the iterative process of the systems development along with the rationale for the choices,  implications and challenges of leveraging LLMs for improving health care in low-resource settings and pathways forward.

\section{Background and Related Work}
We begin by examining the landscape of maternal health in Pakistan, highlighting key challenges and trends (Section ~\ref{subsec:maternal}). We then review emerging applications of AI in healthcare (Section ~\ref{subsec:ai}). Next, we discuss the evolution and impact of EMRs. Finally, we examine the potential of Clinical Decision Support Systems (CDSS) in enhancing patient care and outcomes (Section ~\ref{subsec:cds}).

\subsection{Maternal Health in Pakistan} \label{subsec:maternal}
Pakistan has the world’s fifth-largest population~\cite{pbs2023} and one of the highest maternal mortality ratio in South Asia, at 154 per 100,000 live births~\cite{unwomen_midwife_training, Knoema}. A lack of detailed medical records, limited access to trained health care workers, poor quality services at hospitals, low literacy, and poverty are some of the most significant contributors to maternal deaths~\cite{khan2009maternal, Mustafa2020}. The most significant maternal health complications leading to death are Eclampsia and postpartum hemorrhage~\cite{Shaeen2022}.  Eclampsia, the third leading cause of maternal deaths in Pakistan, is convulsions in a pregnant woman resulting from high blood pressure and is easily manageable if care is provided early on~\cite{firoz2018framework,mir2019expanding}. A significant proportion of maternal deaths in Pakistan can be attributed to delays in providing medical care during obstetric complications. An instance of such delays arises from household constraints, specifically stemming from the ignorance of families and traditional midwives, who postpone the decision to seek necessary medical care in larger hospitals~\cite{Shaeen2022}. Our system aims to address this challenge by raising red flags and highlighting symptoms like those of Eclampsia to ensure care is provided as early as possible. 

 Similarly, anemia during pregnancy is a significant factor in maternal mortality~\cite{noronha2012anemia}, ranking as the second most common cause of such deaths in Asia and contributing to 12.8\% of these fatalities~\cite{khan2006causes}. It is alarming that up to 50\% of pregnant women in Pakistan are estimated to suffer from anemia~\cite{majeed2022anemia}, a condition that can be alleviated with strict monitoring~\cite{rush2000nutrition}. 
 
 The absence of accurate data on maternal mortality and its causes is a major obstacle to reducing maternal deaths in Pakistan \cite{Shaeen2022}. Timely antenatal care is key for detecting and managing pregnancy-related conditions. Essential components of this care include monitoring for hypertension, proteinuria, STIs/HIV, anemia, and fetal malpresentation and educating women and their families about danger signs during pregnancy \cite{Tikmani2019TrendsOA}. The early identification of patients at risk of rapid deterioration and critical illness plays a pivotal role in preventing maternal mortality~\cite{rathore2020validation}. 
 
Our system leverages speech input to generate detailed EMRs for effective monitoring throughout the antenatal period. It offers diagnostic support to healthcare workers by enabling early identification of potential risks, such as anemia, preeclampsia, and gestational diabetes, by generating "red flags" based on the patient's medical history.

\subsection{LLMs in Healthcare\label{subsec:ai}}
In recent years, AI-driven systems have increasingly focused on healthcare, with various workshops exploring the intersection of Human-Computer Interaction (HCI) and AI to address the challenges and opportunities in healthcare environments~\cite{andersen2021realizing, park2019identifying, wilcox2013patient, book}. A key challenge in patient-centered care is the impact of patient-clinician communication, often hindered by time constraints that limit personalized care and clinical decision-making~\cite{burgess2022care, berry2017getting, berry2017creating2, mazaheri2018evaluation, porter2023revisiting}.  To address this, researchers are exploring AI-supported tools for medical treatment and diagnostics, focusing on clinicians' needs to understand an AI models' limitations, perspectives, and design goals to better interpret diagnostic predictions~\cite{Burgess2023, cai2019hello}. This is particularly important in the context of low-resource countries like Pakistan where often the only access to health care is through low-skilled health care workers. 

Recent advancements in LLMs have created significant opportunities in healthcare~\cite{Karabacak2023, Clusmann2023}, for public health monitoring~\cite{10.1145/3544548.3581503, 10.1145/3613904.3642420}, to create structured patient notes and manage clinical notes \cite{cascella2023evaluating, 10.1145/3613905.3650784}, improving patient experiences in hospitals~\cite{10.1145/3613905.3637149}, supporting the mental health of isolated populations~\cite{jo2023conversationalAI}, and working alongside digital voice assistants~\cite{10.1145/3584931.3606993}. Additionally LLM-based applications are currently being piloted to diagnose diseases from clinical notes \cite{jiang2023vetllm}, early detection of health problems \cite{article}, generating medical advice \cite{10.1001/jama.2023.5321}, and supporting decision-making and diagnostics \cite{yu2023generativeAI, tu2024conversationaldiagnosticai, mcduff2023accuratedifferentialdiagnosislarge}. With proper development and supervision, health-focused LLMs are poised to transform various healthcare domains, including consultation, diagnosis, management, and medical education \cite{yang2023large}.

Researchers and health professionals have relied on AI-based tools to address various challenges in maternal health \cite{10.1145/3630106.3658982}. Prominent work in this domain, though limited, includes predicting high-risk pregnancies from Electronic Health Records (EHRs) \cite{Knowles2016HighRP}, aiding community health workers in delivering high-quality maternal care in resource-constrained settings \cite{AlGhadban2023.12.15.23300009}, and analyzing free-speech data about breastfeeding and postpartum depression \cite{inproceedingsOmar23}. An AI-backed tool that captures the complete journey of a pregnant patient's visits, including EMR, and supports clinical decision-making, is yet to exist in the context of a constrained healthcare infrastructure like Pakistan. 

 Much if not all of this work is currently being piloted in the Global North where there is access to extensive resources, skilled health care workers and high adoption of technologies. It is critical that there are parallel attempts to understand the unique potential, constraints and limitations of leveraging LLMs in recourse poor countries where they might be able to improve health outcomes in very specific ways. 

\subsection{Electronic Medical Records} \label{subsec:emr-lit}
Electronic Medical Records (EMR), also known as electronic patient records, facilitate better services and decision-making through accurate medication lists, legible notes, and readily accessible patient data and a record of patient care~\cite{Williams08, Evans2016}. While EMRs are now the standard of care in developed and rich countries, developing countries like Pakistan face significant barriers to adoption. Building successful EMRs for low-resource contexts requires careful consideration of various constraints including a tiered medical system~\cite{Mustafa2020} with healthcare workers ranging from formal to informal~\cite{chen11, MalikKhan09}, digital low-literacy, low proficiency in English, limited funds for expensive hardware, and a lack of training~\cite{DesRoches08,Boonstra2010}. Existing EMRs in resource-constrained regions face a critical limitation due to the lack of customization and scalability~\cite{Ismail2016,usdhhs2009hiv, fraser2005implementing, mamlin2006cooking}. Research addressing the specific requirements of health professionals, who are the primary users of EMR for fetomaternal monitoring, care, and treatment, is still insufficient in Pakistan~\cite{Noor2019}. Researchers developed a prototype for documenting patient history, encompassing past pregnancies, medications, allergies, and surgeries~\cite{Ismail2016}. However, this prototype did not comprehensively address all critical factors pertinent to maternal health, such as socioeconomic history~\cite{Kim2018Socioeconomic}. 
Recent research investigates the integration of LLMs with EHRs, demonstrating how LLMs can support the scrutiny and comprehension of existing EHRs~\cite{Xu2025}, evaluation of smaller LLMs on EHR datasets ~\cite{Qu2024}, and clinical language models designed to enhance EMR documentation~\cite{Yang2022}. However, maternal health in low-resource settings presents unique challenges, where language barriers, infrastructural limitations, and the need for end-to-end deployment remain critical gaps.  

There are existing industry systems that leverage AI to improve clinical documentation for generating EMRs. Leading examples include Epic Systems~\cite{turner2023epic}, which is integrating GPT-4 into EHRs to help clinicians communicate with patients, and Nuance DAX Copilot, which automatically transcribes and summarizes patient encounters \cite{nuance_dax_copilot}. Similar standalone systems like DeepScribe and Chartnote also exist, while others, such as Amazon's HealthScribe and Nabla Co-pilot, require integration with an existing EHR system in place at the hospital \cite{deepscribe, chartnote, aws_healthscribe, nabla_website}. 

While we acknowledge the existence of such systems in developed countries, our approach focuses on addressing the multifaceted challenges encountered in resource-constrained settings. These challenges include varied healthcare worker expertise, digital and English literacy barriers, financial constraints for system hardware, and limited training opportunities. To overcome these hurdles, we have developed a low-cost, smartphone-based AI assistant tailored for Maternal Healthcare Workers (MHCWs) in resource-constrained settings. To the best of our knowledge, this is the first prototype of its kind to be implemented and deployed in a live hospital setting in Pakistan.

\subsection{Clinical Decision Support Systems in Health} \label{subsec:cds}
Clinical decision support  systems (CDSS) play a pivotal role in improving clinicians' decision-making processes by offering tailored insights based on patient-specific data. These systems contribute to various aspects of healthcare, including providing diagnoses, recommending treatments, suggesting preventive care measures, and issuing alerts for potential health concerns \cite{Osherof2007, Greenes2014, Yang2019, Zamora2013}. Working with CDSS poses various challenges, including difficulties integrating them into clinicians' workflows \cite{Maniatopoulos2015}. Poorly designed CDSS can lead to problems such as clinician burnout \cite{Jankovic2020} and system errors \cite{Hartswood2005}. CDSS may be implemented either as independent software applications or as integrated components within EMR systems. Irrespective of their specific configuration, overcoming implementation barriers is critical for ensuring effective adoption and utilization of these systems in clinical environments. While previous studies have explored AI-based decision-support systems \cite{Magrabi2019, Shortlife1975}, the rise of emerging technologies adds a layer of complexity for clinicians to trust these systems. In contrast to rules-based CDSS, where clinicians receive information based on predefined rules, AI-based CDSS can potentially lead to trust issues among clinicians \cite{Duran2021}. Recent studies explore approaches to increase the clinician's trust and adoption of AI-based decision support systems in healthcare \cite{Burgess2023, cai2019hello}. 

In our system, we generate critical health concerns, or "red flags", to aid healthcare workers in delivering prompt interventions. To validate these red flags, we engage senior gynecologists in a thorough evaluation process. This collaborative approach enables continuous model refinement, enhancing the clinical accuracy and relevance of the generated alerts in supporting evidence-based decision-making.

\section{Formative Study: Current Practices and Challenges in Health Data Record Keeping in Hospital Y}\label{subsec:qual}
The Gynecology Department at Hospital Y, a not-for-profit characterized by a high patient influx and limited resources, served as our field site. To ground the system in real clinical practices, we conducted field observations and semi-structured interviews with healthcare professionals. This formative study aimed to gather information on 1) current maternal health data record-keeping practices, challenges in patient management, diagnostic methods, and patients' socioeconomic factors, and 2) the workflow of the Gynae Outpatient Department, and how doctors interacted with patients in real-time.

\subsection{\textbf{Methods}}
We conducted semi-structured interviews with 12 healthcare professionals (Table~\ref{tab:users}). The interview protocol (Appendix~\ref{initialprotocol}) explored current data recording practices, clinical decision-making processes, and sociocultural factors that affect maternal care. Topics included antenatal and postnatal workflows, diagnostic routines, and communication with patients and families.  Two female authors independently conducted contextual inquiries and field observations in the Gynecology Outpatient Department over one week, spending approximately three hours per day on site. Observations covered key clinical spaces, including the waiting area, antenatal and examination rooms, outpatient desks, and consultant offices, with strict attention to patient privacy. Informed consent was obtained from all observed patients. To gain deeper insight into data recording practices, we also conducted participant observation: accompanying a consenting patient throughout her consultation to document real-time interactions and workflows related to patient history recording.
The consultant also provided us with the paper-based template (Antenatal card) for recording patient data (Appendix~\ref{subsec: antenatalcard}) currently being used by the gynae department.

The interviews were transcribed and analyzed using inductive coding~\cite{thomas2006general} by two authors, with codes merged through iterative discussion. Example codes include \textit{`follow-up frequency'}, \textit{`pregnancy detection methods'}, and \textit{`patient education'}. These codes were grouped into themes such as \textit{`inconsistencies in documentation'}, \textit{`informal socioeconomic assessment'}, and \textit{`patient retention challenges'}. 
The field notes independently recorded by the two authors were later combined into a shared account through joint review.
\begin{table}
  \begin{tabular}{clcl}
    \toprule
    Designation & Description & Count & Demographics (/gender) \\
    \midrule
    consultants & \begin{tabular}[t]{@{}p{4cm}@{}}Senior doctors with gynecology specialization\end{tabular} & 2 & > 35 (F) \\
    post-graduate residents & \begin{tabular}[t]{@{}p{4cm}@{}}Doctors formally training in gynecology\end{tabular} & 4 & 25-32 (F) \\
    house officers & \begin{tabular}[t]{@{}p{4cm}@{}}Medical students undergoing rotations at Gynae Outpatient Department\end{tabular} & 4 & 23 - 25 (F) \\
    nurse navigators & \begin{tabular}[t]{@{}p{4cm}@{}}Nursing degree graduates training in gynecology\end{tabular} & 2 & 25 (F) \\
    \bottomrule
  \end{tabular}
  \caption{Participant Composition for Semi-structured Interviews}  
  \label{tab:users}
\end{table}
\subsection{\textbf{Findings of the Formative Study }}
The Gynecology department at Hospital Y manages around 3,000 patients a month. Patient records are entered both manually and electronically, though the system is hampered by time constraints, limited staff availability, and only one computer designated for data entry. When a patient arrives at the hospital, only their essential data (Clinical diagnosis and prescribed medications is entered electronically. However, the detailed information about medical history, family history, and other relevant details, is manually filled out on the hard copy of the antenatal card and handed over to the patient. The key findings of our formative study include:  
\begin{itemize}
    \item \textbf{Paper-based records as the sole patient history:} Patients receive a paper antenatal card that is manually updated by doctors. The card lacks structure and does not contain physical space for long term follow-up. The antenatal card serves as the primary record, with no backup in case the patient loses it or forgets to bring it along during the appointment. 
    \item \textbf{Ad hoc recording of risk factors:} Immediate data such as patient history and high-risk indicators are recorded; doctors use a red pencil to note anything that can potentially complicate the pregnancy.
    \item \textbf{Dependence on patient recall:} Clinicians rely on patients to bring test reports and remember vaccination histories, as no centralized or digital record system is available.  
    \item \textbf{Absence of formal socio-psychological screening:} There is no formal screening for social or psychological issues such as domestic abuse; these are only addressed if the patient voluntarily discloses them. Doctors infer the patient's socioeconomic factors informally through the patient's communication skills and appearance, without taking any formal history.
    \item \textbf{Fragmented patient flow:} After the nurse records the vitals, the patient is randomly assigned to an on-duty doctor. The patient cannot enter the consultant's room without an on-duty doctor, who summarizes the case for the consultant to devise a care plan.
    \item \textbf{Limited digital infrastructure:} Only consultants have access to desktop computers, which are not used for patient record entry. Doctors calculate the gestational age through mobile apps, and the result is manually entered on the antenatal card.  
\end{itemize}

We distilled these findings into initial design requirements (DRs), that directly shaped our system design (Table~\ref{tab:designreqs}). 
\begin{table}[h]
\centering
\begin{tabular}{p{1\linewidth}}
\midrule
\textbf{Design Requirement} \\
\midrule
\textbf{DR1.} Build a structured yet flexible EMR template on mobile that stores longitudinal data across repeat visits and allows quick pre-save edits.\\
\textbf{DR2.} Implement automated red-flag logic that highlights clinically important indicators in the EMR and requires visible acknowledgement or edit before finalizing.\\
\textbf{DR3.} Add a dedicated Socio-Economic History section and targeted prompts so structural constraints are routinely captured.\\
\textbf{DR4.} Add auto-computed “Weeks of Gestation” and “Expected Date of Delivery (EDD)” fields inside the EMR template to eliminate external calculators and reduce error.\\
\textbf{DR5.} Enable non-linear navigation so any section can be completed in any order, with safe partial saves and no workflow breakage.\\
\textbf{DR6.} Extend the EMR with clinically validated fields and LLM-prompted follow-ups that surface commonly missed symptoms during capture.\\
\textbf{DR7.} Add an image-upload feature for lab/scan reports so clinicians can attach artifacts directly to the visit record.\\
\bottomrule
\end{tabular}
\caption{Design requirements derived from formative study and iterative testing.}
\label{tab:designreqs}
\end{table}\\
These design requirements formed the foundation for our system design, guiding the translation of observed challenges into concrete features and workflows (Section ~\ref{sec:system2}).

\section{System Design and Development}
\label{sec:system2}
\subsection{System Overview}
Our application is designed as a modular mobile voice-based AI system to support maternal healthcare documentation in low-resource clinical environments. This section outlines the system architecture, interface design, and technical development of the ASR and LLM modules, as well as the iterative testing processes that shaped the final deployment. The system consists of three key modules:
\begin{enumerate} \item A structured mobile interface that allows non-linear navigation and speech-based data entry. \item An ASR module, based on OpenAI's Whisper, fine-tuned on locally collected Roman Urdu (Urdu written using the Latin script) speech data. \item A GPT-4 LLM backend that processes transcriptions to generate EMRs, clarification questions, and identify clinical red flags. \end{enumerate}
Data flows from the user through speech input, which is transcribed by the ASR and then passed to the LLM with medical dictionaries and prompt templates. The resulting EMRs and red flags are displayed in the app for review. In Section~\ref{subsec:ui}, we describe the user interface and core features for recording patient histories and generating EMRs. We then present the technical details of our text generation and speech recognition modules in Section~\ref{subsec:text} and Section~\ref{subsec:speech}.

\begin{figure}[ht]
\centering
  \includegraphics[width=\linewidth]{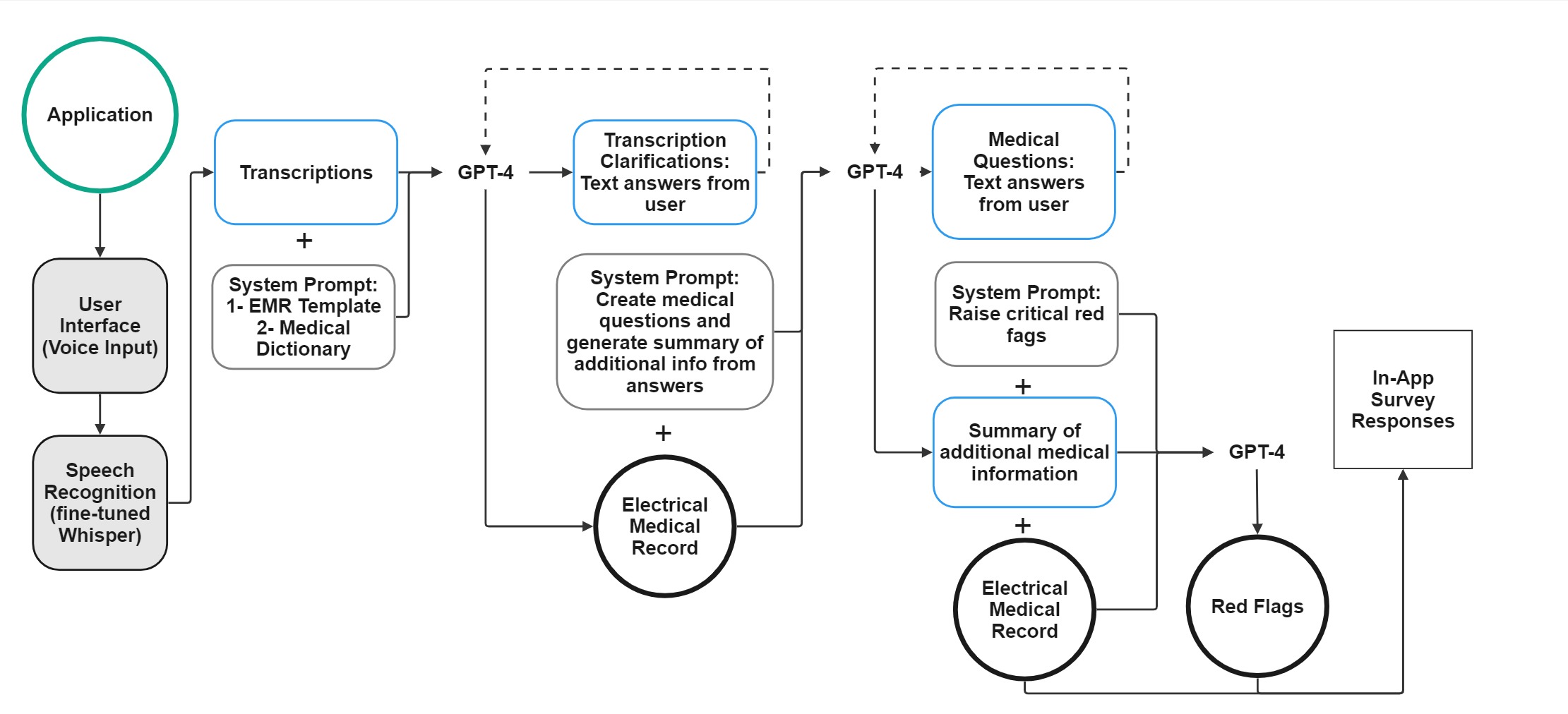}
  \caption{An overview of our system}
  \Description{A flow chart diagram showing an overview of the system}
   \label{fig:system-design}
\end{figure}

\subsection{\textbf{Scope of Deployment}}  
The broader vision for System~X extends beyond physicians to include maternal healthcare workers who constitute the backbone of frontline care in Pakistan. Nationally, 31\% of deliveries are still attended by unskilled providers~\cite{nips2019pdhs}, such as traditional birth attendants (\emph{dais}), majority of whom are unable to read or write in either Urdu or English~\cite{Mcnojia2020}. This underscore both the reach of formal care and the persistent reliance on low-literacy providers in peripheral settings.  
Deploying LLMs in such high-stakes domains carries inherent risks, particularly the potential for EMR or decision-support errors that could directly compromise patient safety.  Unlike doctors, unskilled workers cannot independently verify the accuracy of outputs or compensate for model errors. For this reason, we first piloted the system with doctors and nurses to ensure it performed as intended before considering broader use. Our long-term goal remains to adapt and extend the system for maternal healthcare workers in low-resource, low-literacy environments, where voice input and structured EMRs could substantially improve record-keeping, continuity of care, and patient safety.  

\subsection{User Workflow and Interface}
\label{subsec:ui}

\begin{figure*}[t]
\centering
\begin{minipage}[t]{0.32\textwidth}
  \centering
  \includegraphics[height=0.80\textheight]{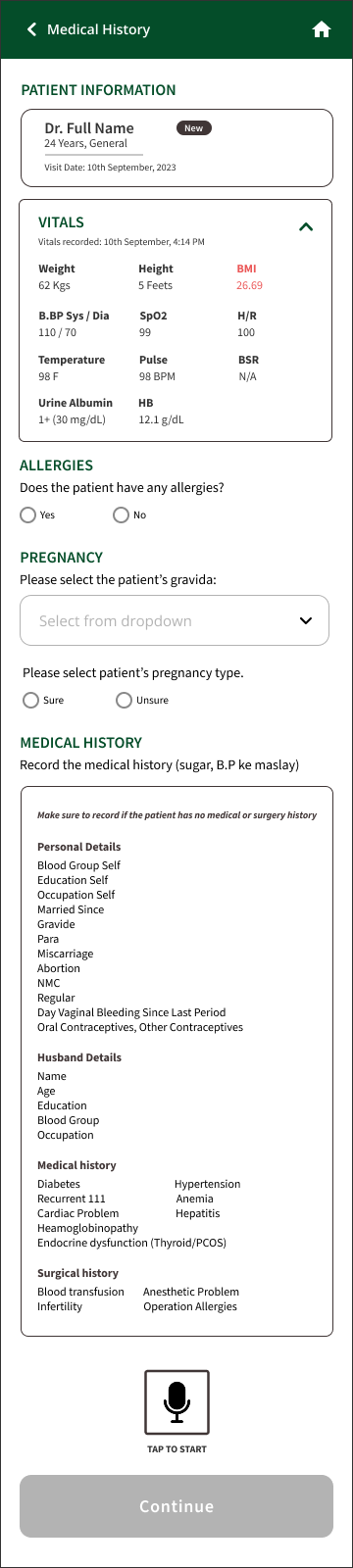}
  \captionof{figure}{Expanded and scrollable "Medical History" screen shown in full for illustration. In the actual application, only a portion of the screen is visible at a time. Clinicians tap the microphone icon to record patient history in their preferred language.}
  \label{fig:history-collapsed}
\end{minipage}
\hfill
\begin{minipage}[t]{0.32\textwidth}
  \centering
  \includegraphics[height=0.80\textheight]{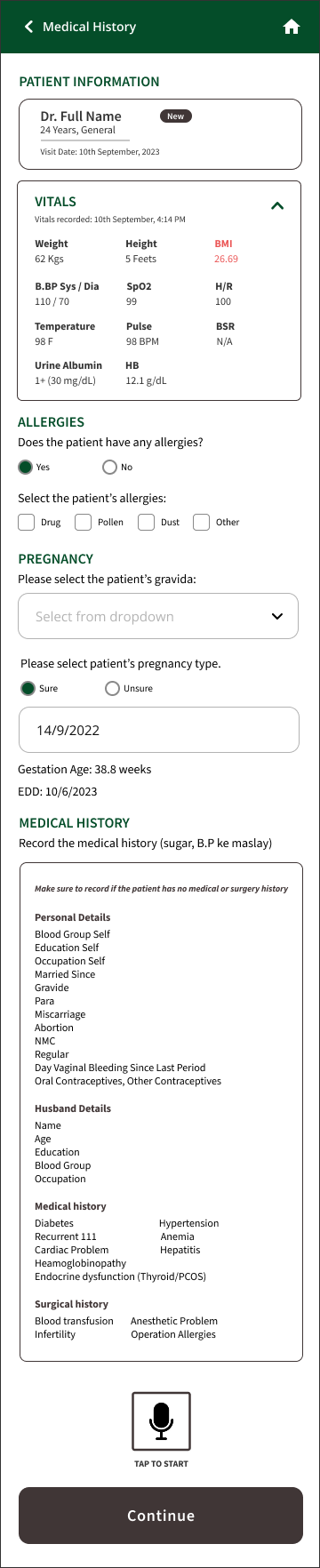}
  \captionof{figure}{Filled and scrollable "Medical History" screen displaying recorded allergies and past medical conditions. The combination of text fields and audio input reflects design decisions shaped by formative research and iterative testing.}
  \label{fig:history-filled}
\end{minipage}
\hfill
\begin{minipage}[t]{0.32\textwidth}
  \centering
  \includegraphics[height=0.80\textheight]{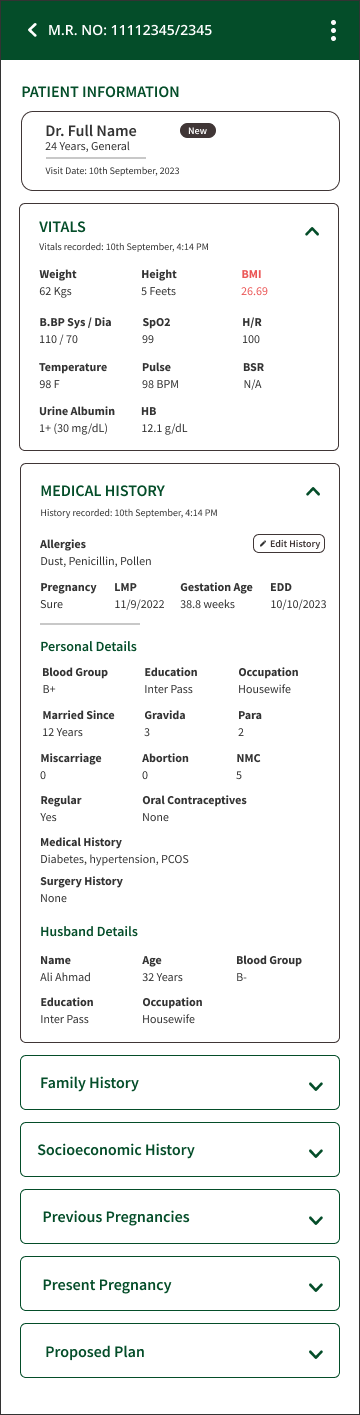}
  \captionof{figure}{Scrollable patient summary screen displaying vital signs and collapsible history categories. The MR No. at the top serves as the patient’s unique medical identifier.}
  \label{fig:patient-info}
\end{minipage}
\end{figure*}

Our smartphone-based system guides healthcare workers through a structured workflow for EMR generation using speech input. The process begins with basic patient data entry (Section~\ref{subsubsec:Patient Information Entry}), followed by voice-recorded history across six categories (Section~\ref{subsubsec: Recording and Structuring}). Transcriptions are reviewed and clarified through LLM-generated questions (Section \ref{subsubsec:step3}), leading to the finalized EMR (Section ~\ref{subsubsec:step4}). Additional medical questions (Section ~\ref{subsubsec:step5}), followed by red flag alerts (Section~\ref{subsubsec:red}) and ultrasound uploads (Section ~\ref{subsubsec:step7}) further enhance clinical decision-making. The overall data flow and system outputs are illustrated in Figure~\ref{fig:system-design}.

\subsubsection{\textbf{Step 1: Patient Information Entry}}
\label{subsubsec:Patient Information Entry}
In our system workflow, nurses first input basic patient information, including name, age, healthcare type (public or private), and vital signs such as height, weight, blood pressure, temperature, and pulse. Doctors then retrieve the patient record via a unique medical ID and record patient information. Figure~\ref{fig:history-collapsed} illustrates how the recorded vital signs are displayed after being entered by the nurse. Assigning a unique medical ID enables longitudinal retrieval across visits and avoids repeated re-entry, directly addressing record fragmentation \textbf{[DR1]}.

\subsubsection{\textbf{{Step 2: Recording and Structuring Patient's Information}}}
\label{subsubsec: Recording and Structuring}

After retrieving the patient record using a unique medical ID, doctors record the patient’s history using the in-app audio recorder, speaking in Urdu, English, or a mix of both.  
Informed by insights from the formative study, the application structures patient history into six key sections: (1) Personal and Medical History, (2) Family History, (3) Socio-Economic History, (4) Previous Pregnancy Details, (5) Current Pregnancy Details, and (6) Follow-up Plan. Each category is presented on a separate screen. For returning patients, only categories (5) and (6) are updated \textbf{[DR1]}. Speech is recorded in Urdu, English, or code-switched combinations. Figure~\ref{fig:history-collapsed} shows the unfilled "Personal and Medical History" screen, while Figure~\ref{fig:history-filled} displays the corresponding section after the doctor has recorded the data. Figure~\ref{fig:patient-info} displays all six collapsible patient information sections.

To accommodate diverse documentation practices, the interface supports non-linear navigation, allowing clinicians to complete history sections in any order. Field observations revealed that while some follow the antenatal card sequence, others organize information based on clinical reasoning. This design flexibility aligns with existing workflows, minimizing disruption and enhancing usability [\textbf{DR5}]. We also include in-template auto-computed \textit{`Weeks of Gestation} and \textit{`Estimated Date of Delivery (EDD)'} eliminating external calculator apps and transcription steps \textbf{[DR4]}.

Through iterative design and testing, additional clinically relevant data fields not present on the physical antenatal card were identified and incorporated, such as burning micturition, PV discharge, vaginal itching, and gastrointestinal symptoms. These additions were validated by a senior gynecologist, resulting in a more comprehensive EMR template tailored to local diagnostic practices \textbf{[DR6]}.

\subsubsection{\textbf{Step 3: Transcription Review and Clarification}}
\label{subsubsec:step3}
After completing the recordings, the doctors are presented with a screen (Figure \ref{Clarifications}) displaying the transcribed text for each category. Similar to the recording screens, the transcription screens are separate for each category. In these screens, the transcription is presented alongside a list of clarification questions generated by the LLM to ensure accurate EMR creation. These questions typically address misspelled words and confirmation of the doctor's spoken content, requiring simple answers shown on screen (Figure \ref{Clarifications}). Users provide text input to respond to these questions, and the process follows a linear flow to prevent any questions from being skipped.
\subsubsection{\textbf{Step 4: EMR Review and Finalization}}
\label{subsubsec:step4}
Once clarifications are submitted, a consolidated EMR is generated and presented on a single, scrollable screen. Each category can be expanded or collapsed for easier review. Doctors may edit any field or add supplementary notes before saving \textbf{[DR1]}. At this stage, an in-app survey also prompts feedback on EMR accuracy and completeness.

\subsubsection{\textbf{Step 5: Supplementary Medical Questions}}
\label{subsubsec:step5}
Following the review of the EMR, users are prompted to respond to a set of additional medical questions generated by the system that aims to capture information not in the EMR via text input, e.g., \textit{"As the patient in a consanguineous marriage, are there any inherited diseases in their families that should be monitored?"}  \textbf{[DR6]}. See Appendix~\ref{subsec:medquestions} for further examples of these questions. 

\subsubsection{\textbf{Step 6: Red Flag Review}}
\label{subsubsec:red}
In the final stage of the workflow, MHCWs review critical red flags generated by the system. These red flags are highlighted in red text in the EMR, mirroring existing paper-card markups, and commonly include issues such as gestational diabetes, anemia, or abnormal body mass index  \textbf{[DR2]}. Missing information in the patient’s history, such as absent vitals, lab tests, or gaps in medical/surgical history, is highlighted in yellow text. Finally, an in-app survey is displayed to capture the doctor’s feedback on the usefulness of the red flags.

\subsubsection{\textbf{Step 7: Ultrasound Report Integration}}
\label{subsubsec:step7}
For returning patients, the application includes the additional feature of uploading an image of the ultrasound scan lab report {\textbf{[DR1]}. Upon upload, the system extracts key information from the image, such as fetal movement, placenta presence, date of the scan, and the presence of anomalies. This streamlined process minimizes the time required for doctors to manually read the entire report, allowing them to focus directly on the EMR and make edits if necessary.

\subsection{Fine-Tuned Speech Recognition System}\label{subsec:speech}
\subsubsection{Rationale for Speech as the Primary Input Modality}
Our choice of speech as the primary input modality stems directly from the realities of maternal healthcare in Pakistan. A significant proportion of births (31\%) are still attended by unskilled providers~\cite{nips2019pdhs} cannot read or write in Urdu or English~\cite{Mcnojia2020}. In such contexts, requiring text-based input would exclude a large segment of frontline workers. In contrast, speech input lowers barriers to adoption. Hands-free, phone-based dictation allows providers to capture data without the cost and space demands of workstation setups, while supporting natural interaction in Urdu or code-switched Urdu/English. 

Prior work shows that globally physicians spend nearly twice as much time documenting as providing care~\cite{Sinsky2016}. Speech recognition in EHRs can reduce turnaround times, enhance report completeness~\cite{Saxena2018, Goss2019, KumahCrystal2018}. Hybrid approaches increase documentation speed by 26\% while doubling document length~\cite{KumahCrystal2018}. These efficiencies are particularly valuable in high-volume, resource-limited clinics.

At Hospital Y, only senior consultants have access to computers, which are not used for data entry. The situation is even worse in smaller hospitals and tertiary health units where such infrastructure is often absent altogether. Hands-free, phone-based dictation lets healthcare workers capture patient data, avoiding the space and cost overhead of workstations. Speaking in Urdu (or mixed Urdu/English) also lowers the entry barrier for junior staff. We counter noisy outpatient environments via (i) a custom medical dictionary, and (ii) an automatic clarification loop, to reduce transcription errors propagating to the final EMR or red-flag logic.
Thus, voice input is a context-aware design choice that addresses digital literacy, infrastructural limitations, and documentation bottlenecks.

\subsubsection{Model Selction and Fine-tuning}

We used the \texttt{large-v2} version of OpenAI’s Whisper model as the base speech recognition system for transcribing patient interactions. To assess its baseline performance, we collected a pilot dataset comprising audio recordings from 8 patients (approximately 16 minutes of speech). These recordings reflected typical outpatient consultations, including frequent code-switching between Urdu and English. We benchmarked Whisper against the MMS-1b-all model~\cite{mms} (default configuration). As shown in Table~\ref{tab:WER_Comparison}, both models struggled with code-switched input: Whisper achieved a Word Error Rate (WER) of 45.71\%, while MMS-1b-all recorded 60.89\%. WER denotes the proportion of words incorrectly transcribed.

Given these high error rates, we fine-tuned Whisper using pilot dataset. Because Whisper requires training samples under 30 seconds, we segmented the recordings using Praat~\cite{praat}. Each segment was manually transcribed into two formats: (i) Roman Urdu and (ii) Arabic-script Urdu (with transliterated English terms).
Initial full-model fine-tuning degraded performance, likely due to overfitting on the small dataset. To address this, we adopted a parameter-efficient strategy using Low-Rank Adaptation (PEFT-LoRA)~\cite{fastwhisper_finetuning}, restricting trainable parameters to 1\% of the model. This approach improved accuracy substantially: WER decreased to 16.38\% for Latin Urdu and 28.85\% for Arabic-script Urdu.

For deployment, we integrated the fine-tuned Latin Urdu model with \texttt{faster-whisper}~\cite{fasterwhisper}, a lightweight decoding implementation optimized for real-time inference. The final model was hosted on a Hugging Face inference endpoint and connected to the mobile application for on-device transcription. Table~\ref{tab:WER_Comparison} summarizes baseline and fine-tuned performance.

We further mitigate challenges of noisy outpatient environments by (i) incorporating a custom medical dictionary to resolve local terminology (Section~\ref{subsubsection:medDictionary}) and (ii) embedding an automatic clarification loop that prompts clinicians to confirm uncertain transcriptions (Section~\ref{subsubsection:clarificationquestions}). Together, these features help prevent transcription errors from propagating into the final EMR or compromising the red flag alerts.
\begin{table}[htbp]
    \centering
    \begin{tabular}{|p{0.4\textwidth}|p{0.2\textwidth}|}
    \hline
    Model & WER \\
    \hline
    MMS-1b-all & 60.89\% \\
    Whisper-large-v2-base & 45.71\% \\
    Whisper-large-v2-fine-tuned-urdu & 28.85\% \\
    Whisper-large-v2-fine-tuned-roman-urdu & 16.38\% \\
    \hline
    \end{tabular}
    \caption{Comparison of Word Error Rate(WER) for different ASR Models}
    \label{tab:WER_Comparison}
\end{table}

\subsection{LLM Integration for EMR and Diagnostic Support}
\label{subsec:text}
Our system utilizes OpenAI's GPT-4 model for various functions, including generating EMRs, prompting doctors for medical questions, and detecting potential health risks or `red flags' in the generated EMRs.

\subsubsection{\textbf{Iterative Improvements in EMR Generation Workflow}}
\label{subsubsection:clarificationquestions}
Initially, the LLM populated structured EMR templates directly from ASR transcriptions using predefined clinical parameters. However, early tests revealed limitations: fields omitted by doctors led to vague outputs (“No information”), off-template details were discarded, and Roman Urdu transcription errors were often preserved. To address this, we revised the prompt design by adding an "Additional Information" field and instructing the model to attempt basic error correction. We also introduced a clarification step where the LLM generates follow-up questions for ambiguous inputs (Figure~\ref{Clarifications}), enabling doctors to disambiguate via brief text responses. We instructed doctors to explicitly say “no information” when a field was not applicable, and configured the model to default to “No” for parameters not explicitly mentioned. These changes significantly improved data completeness and EMR accuracy while preserving clinical nuance.

\subsubsection{\textbf{Medical Dictionary Integration}}
\label{subsubsection:medDictionary}
To reduce the volume of clarification questions, the model was supplemented with a medical dictionary. This dictionary was created by comparing system-generated transcriptions with manually generated ground truth transcriptions of patient data recordings, validated by senior gynecology and obstetrics consultants at Hospital Y . The medical dictionary, patient data transcriptions, and EMR template are provided to the model as inputs, and the model is prompted to generate 'clarification questions' for these inputs. The answers to these questions are fed back into the model to produce the final generated EMR. Figures ~\ref{Clarifications} and \ref{EMR} illustrate an example of the clarification questions posed and the EMR generated, respectively, for a sample `Proposed Plan' recording. Appendix \ref{subsec:med-vocab} lists the medical dictionary supplied to the model.

\begin{figure*}
\centering
\begin{minipage}[b]{0.45\textwidth}
\includegraphics[width=1.0\linewidth]{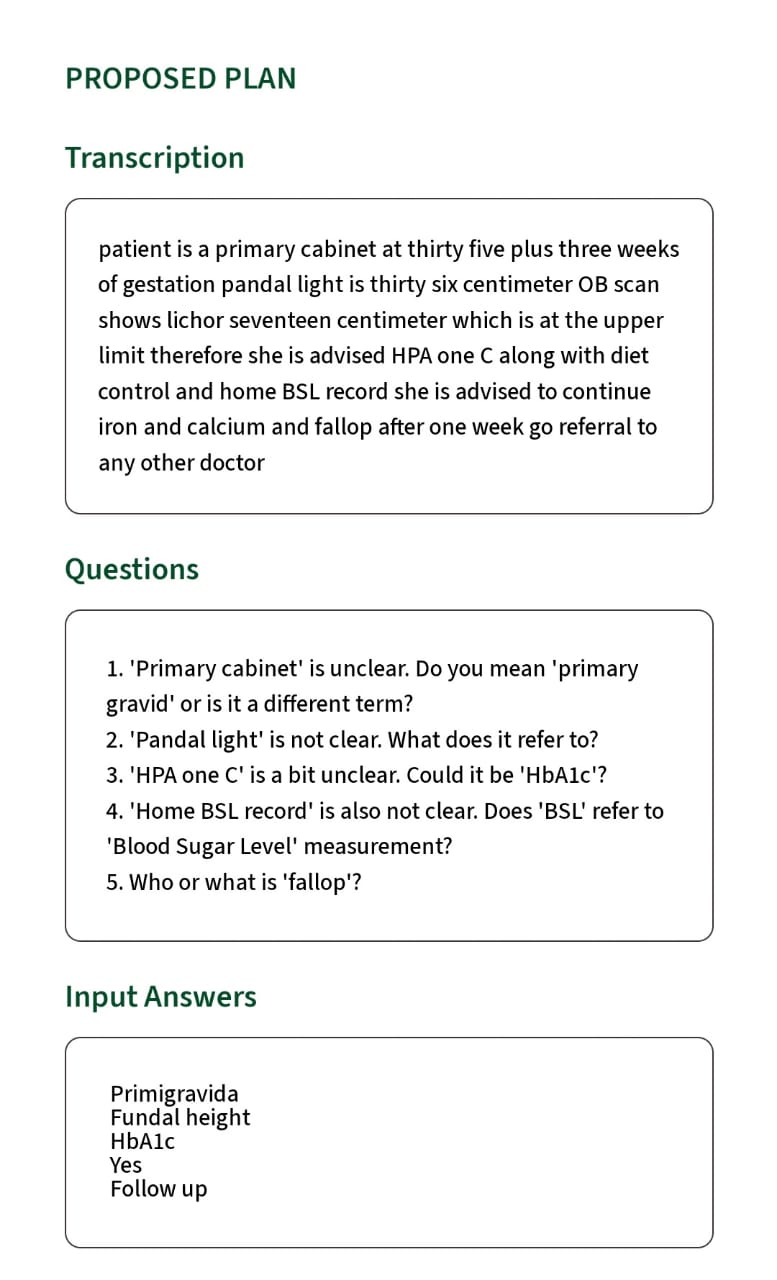}
\caption{Sample Clarification Questions Screen of a Patient's Proposed Plan}
\Description{A sample clarification questions screen of a patient's Proposed Plan}
\label{Clarifications}
\end{minipage}\hfill
\begin{minipage}[b]{.45\textwidth}
\includegraphics[width=1.0\linewidth]{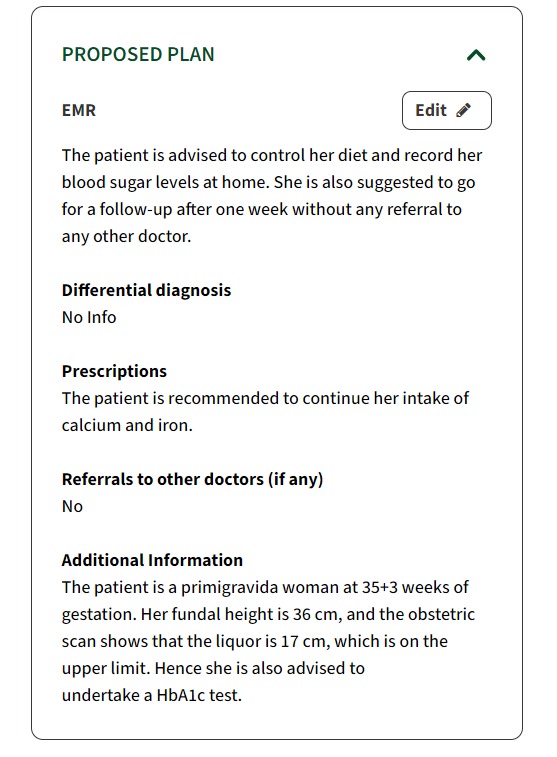}
\caption{Sample EMR screen of a patient's `Proposed Plan'}
\Description{A sample EMR screen of a patient's proposed plan}
\label{EMR}
\end{minipage}
\end{figure*}

\subsubsection{\textbf{Generating Medical Questions}}
The purpose of these questions is to collect additional patient information that might be valuable but is not included in our EMRs (See Appendix~\ref{subsec:medquestions} for examples of these questions). The LLM generates these questions using the previously created patient EMR as a basis to explore additional dimensions of the patient's health status. The LLM is instructed in the prompt to summarize any new information obtained from the responses to these questions.

\subsubsection{\textbf{Red Flags Generation}}
Following consultation with a senior gynecologist, we refined the system’s risk detection logic to prioritize clinically critical pregnancy indicators. Thresholds were informed by guidelines from the Royal College of Obstetricians and Gynaecologists (RCOG)~\cite{rcog} and incorporated directly into the LLM prompt template. The following parameters were designated as high-alert indicators(Table~\ref{tab:clinical-thresholds}).:

\begin{table}[h!]
\centering
\begin{tabular}{|p{6cm}|p{6cm}|}
\hline
\textbf{Clinical Parameter} & \textbf{Threshold} \\ \hline
Blood Pressure & $> 140/90$ mmHg \\ \hline
Body Mass Index (BMI) & $> 30$ kg/m$^2$ \\ \hline
Hemoglobin & $< 11$ g/dL \\ \hline
Random Blood Glucose or HbA1C & $\geq 160$ mg/dL or $\geq 7\%$ \\ \hline
Urine Dipstick & Albumin $\geq$ 1+, Glucose $\geq$ 2+ \\ \hline
\end{tabular}
\caption{Clinical Parameters and Thresholds for Red Flag Generation}
\label{tab:clinical-thresholds}
\end{table}

The LLM is instructed to use these thresholds to flag potential maternal health risks and identify missing patient information. The output is organized into two categories: \textbf{Critical Red Flags} and \textbf{Missing Information}.

Keeping the generated content concise and easy to read was a major challenge. To address this, we modified the model's instructions, emphasizing the need for concise output tailored for obstetrician review. Despite these adjustments, our ongoing focus remains on achieving a balance between content reduction and maintaining clarity and comprehensiveness.

\section{Deployment and Study Design}
\label{sec:deployment}

We deployed our system in the Gynecology Outpatient Department of a large non-profit hospital in Pakistan from October 2023 to April 2024. IRB approval was obtained from Hospital Y, and informed consent was collected from both healthcare workers and patients. The system was piloted in parallel with existing paper-based workflows, ensuring no disruption to routine clinical care. The system was used to supplement documentation and highlight clinical risks, and participating doctors retained full control over clinical decisions and received compensation for their additional time.

A total of 26 maternal healthcare workers - including consultants, residents, house officers, and nurse navigators - used the system during the pilot, generating more than 500 real-time EMRs. All data were anonymized and securely stored. In line with the European Commission’s Ethics Guidelines for Trustworthy AI~\cite{EuropeanCommission2019}, the deployment prioritized human oversight, transparency, and fairness. Continuous feedback through in-app surveys, LLM interactions, and clinician interviews informed iterative improvements and supported safe, inclusive integration of AI tools in low-resource healthcare settings.

\section{Evaluation and Findings}
\label{sec:evaluation}

Our field deployment included both quantitative and qualitative methods to evaluate the effectiveness of key components, including the user interface, Whisper-generated transcriptions, and system-generated EMRs and red flags.  
During initial piloting , we  collected data to fine-tune Whisper and improve prompts. This initial data is not included in our evaluation. We also removed 19 incomplete patient records. Therefore, our analysis is based on the remaining 297 patient records. In this section, we report our findings based on:
\begin{itemize}
\item Usability testing of application (Section~\ref{subsec:usabilitytesting})
\item Evaluation of the speech-to-text module using WER analysis (Section~\ref{subsec:speech_wer})
\item Accuracy measurement of LLM-generated EMRs by calculating correctly labeled fields (Section~\ref{subsec:accuracyEMRs})
\item Categorization of inaccuracies within the LLM-generated EMRs by gynecologist (Section~\ref{subsec:categorization})
\item Assessment of the medical accuracy and relevance of system-generated red flags by gynecologists (Section~\ref{subsec: RFvalidation})
\item Endline semi-structured feedback interviews with the healthcare team (Section~\ref{subsec: feedbackinterviews})
\end{itemize}
\begin{figure}[htb]
\centering
  \includegraphics[width=\textwidth]{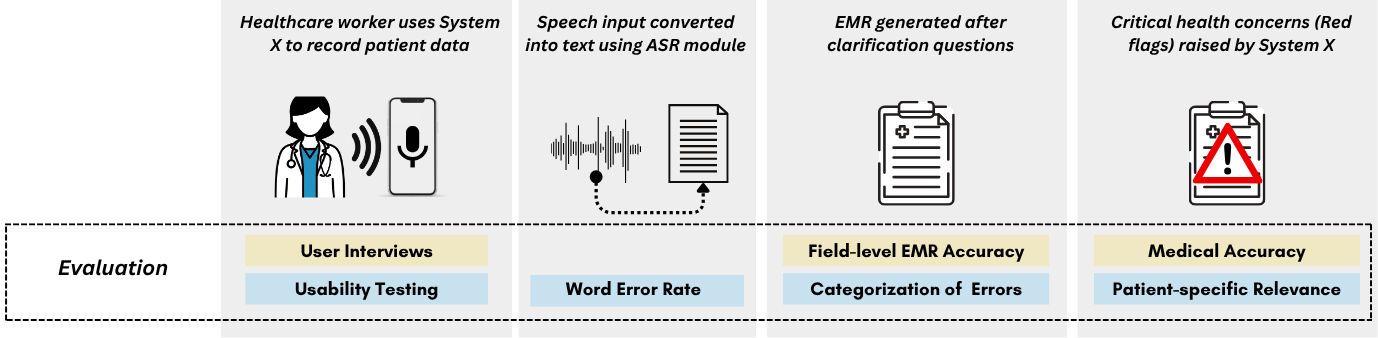}
  \caption{Overview of the system evaluation}
  \Description{Overview of the system evaluation}
   \label{fig:system-analysis}
\end{figure}

\subsection{Usability Testing}
\label{subsec:usabilitytesting}
\subsubsection{Methods: Participants and Procedure.} Usability testing was conducted with seven randomly selected healthcare professionals from the gynecology department, none of whom had prior experience using smartphone applications for patient record management. 
All tests were conducted on-site at the gynecology outpatient clinic of Hospital Y, using smartphones provided by our research team. This phase focused on evaluating the application's usability by applying specific metrics, such as task completion time, the frequency of help requests, and the number of incorrect clicks during task execution \cite{nngroup2024}. Additionally, it aimed to identify ways to improve the user interface for patient data collection and gather feedback on specific features and overall navigation. Participants received an introduction to the application and were informed that they would perform specific tasks independently. They were encouraged to think aloud during the process. Two facilitators guided the sessions, providing explanations of the scenarios and tasks, listed in Table~\ref{tab:usability-results}. 

Informed consent was obtained from patients whose clinical data were recorded in the application. Additionally, consent for video and screen recording was taken from the participating healthcare professionals as they executed the assigned tasks. This was necessary for verifying the accuracy of tester notes and recorded metrics. At the end of each test, we administered a System Usability Scale (SUS) questionnaire \cite{brooke1996sus}. 
\subsubsection{Findings: Easy-to-use Application} 
The SUS score averaged 76, indicating good usability, with individual scores ranging from 65 (above average) to 85 (excellent). The overall task completion rate was 91.44\%, and participants generally found the application easy to use. Table~\ref{tab:usability-results} presents the detailed task breakdown.

\begin{table}[h]
\centering
\resizebox{\textwidth}{!}{%
\begin{tabular}{|l|p{1.8cm}|p{1.5cm}|p{1.7cm}|p{1.9cm}|}
\hline
\textbf{Task} & \textbf{Completion Rate (\%)} & \textbf{Avg Time (s)} & \textbf{Avg Help Requests} & \textbf{Avg Incorrect Clicks} \\ \hline
1. Account Creation & 100 & 59.2 & 0.67 & 1 \\ \hline
2. Search for Patient & 100 & 18.7 & 0.14 & 0 \\ \hline
3. Fill \& Record Medical History & 85.7 & 153 & 1.2 & 0.3 \\ \hline
4. Navigate to Proposed Plan Screen & 100 & 13.3 & 0.25 & 0 \\ \hline
5. Review Transcriptions \& Answer Clarifying Questions & 100 & 103 & 0 & 0.5 \\ \hline
6. Review/Edit Generated EMR & 100 & 63.5 & 0.5 & 0 \\ \hline
7. Answer Additional Medical Questions & 50 & 203.5 & 1 & 0 \\ \hline
8. Review Red Flags & 100 & 55.7 & 0.6 & 0 \\ \hline
\textbf{Average} & \textbf{91.44} & \textbf{83.61} & \textbf{0.54} & \textbf{0.23} \\ \hline
\end{tabular}%
}
\caption{Usability Testing Results}
\label{tab:usability-results}
\end{table}
 
 Although the scores were high, the participants still needed initial training to understand the concept of the application. There were few help requests and incorrect clicks on average, but the variation between sections using only speech input and those combining speech with traditional input methods (e.g., date pickers) led users to overlook non-speech fields.  For example, in task 3, the only unsuccessful attempt occurred when a doctor missed entering the LMP (Last Menstrual Period) date because the field lacked a clear label. To resolve this problem, we improved the visibility of the field by changing its label from \textit{LMP} to more explicit \textit{Please enter the patient's LMP date}.
Navigation was generally smooth, with users correctly identifying the \textit{Continue} button to move forward and using the back arrow in the header to return to previous screens. The testing identified minor improvements, such as removing redundant screens, consolidating patient history into a single dropdown, and refining prompts to reduce repetitive questions, enhancing the application's usability for future iterations.

\subsection{Effectiveness of Speech-to-Text Model}
\label{subsec:speech_wer}
Feedback from clinical collaborators highlighted the need for more targeted evaluation metrics, particularly regarding transcription accuracy. Due to the sensitivity of medical data and privacy concerns, all transcriptions were handled internally. Three members of the research team manually transcribed audio recordings for 297 patients, creating the ground truth dataset for evaluating the speech-to-text module.

We computed Word Error Rate (WER) by comparing system-generated transcriptions with these manual references, where a lower WER indicates better performance~\cite{MicrosoftSpeech2024}. The fine-tuned Whisper model initially produced a WER of 32\%. Upon manual inspection, we observed repetitive phrase generation—likely a result of Whisper’s sequence-to-sequence architecture~\cite{OpenAIWhisper2024}. For example, in the case of patient P302, the correct transcription was "surgical history is not significant." However, Whisper incorrectly transcribed it with excessive repetition: "surgical history is not significant surgical history is not significant... surgical history is not significant". Although these repetitions did not cause errors in the system-generated EMRs, they significantly increased the WER. After removing the repetitive text, the WER was reduced to \textbf{16.2\%}, falling within the acceptable range\cite{OpenAIWhisper2024}. 

The Urdu language uses a modified Perso-Arabic script. When transliterated into the Roman script, this leads to spelling variations due to the lack of standardization in Roman Urdu. For example, a single Urdu word can be written in Roman script as \textit{`kertay', `krte', `krtay',} or \textit{`krty'}, all meaning \textit{`do'} in English. In WER calculations, these variations between the manual and Whisper transcriptions are treated as distinct words, potentially increasing the WER percentage.

\subsection{Accuracy of System-Generated EMRs}
\label{subsec:accuracyEMRs}
For each patient, we used the manually generated `ground truth' transcriptions to create a `ground truth EMR'. Three members of our research team created these ground-truth EMRs by manually filling the EMR template using the ground-truth transcriptions. 
We then compared system-generated EMRs and ground-truth EMRs field-by-field and counted each mismatch in two corresponding fields as an error. Errors in the system-generated EMRs included misspelled medicine names and incorrectly marking a field as `Not mentioned' instead of `No', etc. We quantified the accuracy of the EMRs by calculating the ratio of correctly filled fields by our system to the total number of fields in each EMR. The
evaluation resulted in a system accuracy of \textbf{96.2\%}. The manual transcription of audio recordings and manual generation of the ground-truth EMRs enabled us to differentiate between inaccuracies introduced by the language model and those carried over from the Whisper transcriptions, helping us identify areas for improvement in both modules.

\subsection{Categorization of Inaccuracies in System-Generated EMRs}
\label{subsec:categorization}
After calculating the overall accuracy of system-generated EMRs, we analyze the incorrectly filled EMR fields with the help of a senior consultant gynecologist. 
Inaccuracies in the EMR can lead to poor decision making and incorrect follow-up care plans.  We observed how healthcare workers handle these inaccuracies and, in consultation with a senior gynecologist, adapted the framework from~\cite{unknown23} to create three categories for incorrect fields: \textit{No Action Needed}, \textit{Easily Identifiable and Correctable}, and \textit{Unidentifiable and Uncorrectable Without Ground Truth}. 

We provided the consultant with an anonymized Excel sheet displaying system-generated and ground-truth EMRs side by side. Inaccurate fields were highlighted in red, and correct fields in green, and the consultant was asked to categorize each error. Table \ref{emr-table} summarizes the consultant's categorization of inaccuracies identified in the EMRs, along with their corresponding percentages of distribution. These inaccuracies mainly included misspellings, missing information, or misclassified fields. For example, a misspelled drug name was typically labeled as \textit{Easily Identifiable and Correctable} or \textit{Unidentifiable and Uncorrectable Without Ground Truth}, whereas a misspelling in the patient’s education level was often considered to require \textit{ No action needed}. The consultant perceived errors in clinically significant fields, such as the number of previous abortions or miscarriages, as more consequential than inaccuracies in fields such as gravida or the birthplace of previous deliveries. Table \ref{emr-table} suggests 97.5\% of the EMR fields are actionable, with <2.5\% requiring ground truth for correction. Nonetheless, both \textit{Easily Identifiable and Correctable} and \textit{Unidentifiable and Uncorrectable Without Ground Truth} errors, emphasize the importance of improving EMR accuracy, particularly by reducing the frequency of inaccuracies that are difficult to detect or verify without access to ground truth.

\begin{table}[h!]
\renewcommand{\arraystretch}{1.2}
\setlength{\tabcolsep}{4pt}
\centering
\begin{tabular}{|p{3.2cm}|p{1.2cm}|p{4cm}|p{6.6cm}|}
\hline
\textbf{Category} & \textbf{\%} & \textbf{Definition} & \textbf{Common Examples} \\
\hline
No Action Needed & <1\% & The inaccurate EMR field is insignificant and does not impact clinical decision-making, so it is disregarded. & {\raggedright Spelling mistake in place of birth in `previous pregnancies'.\\ Patient's education marked as `Metric' instead of `Matric'.\\ Gravida marked `Prime (First Pregnancy)' instead of `Primigravida (First Pregnancy)'.\par} \\
\hline
Easily Identifiable and Correctable & <1\% & The doctor can easily spot the inaccuracy by reviewing the patient’s complete EMR and make the necessary correction without additional input. & {\raggedright Incorrect medicine name (Duplascon instead of Duphaston).\\ `Vaginal bleeding since last period' marked `yes since week ago' instead of `since 23rd August 2023'.\\ Family status of patient marked unknown instead of nuclear family.\par} \\
\hline
Unidentifiable and Uncorrectable Without Ground Truth & <2.5\% & The errors cannot be identified or corrected by the doctors without access to the original recording or an external ground-truth EMR. & {\raggedright Age of baby in `previous pregnancy details' marked as 3 years instead of `No Info'.\\ Spelling mistake in husband's name.\\ `Abortions' marked as 2 instead of 1.\par} \\
\hline
\end{tabular}
\caption{Categorization of EMR Inaccuracies by Consultant Gynecologist}
\label{emr-table}
\end{table}

\subsection{Clinical Validation for Red Flags}
\label{subsec: RFvalidation}
Our system includes a `red flag' feature that alerts healthcare professionals to anomalies or critical issues in patient data providing diagnostic support. This feature is designed to facilitate early detection of potential issues, allowing healthcare professionals to implement proactive interventions. In our evaluation of the red flag feature, we found no existing framework in the literature that provided specific metrics for its assessment. Consequently, following collaborative discussions within the team, we developed an approach centered on ensuring the medical accuracy of the red flags. However, during open-ended discussions with doctors using the system, they suggested refining the red flags to be more specific, providing clearer indications of issues related to each condition, rather than their current, medically accurate but generalized nature. This feedback highlighted the need to revise our evaluation metrics. In response, we collaborated with senior gynecologists to develop a more targeted evaluation metric. We modified the system based on their recommendations and evaluated the quality of identified red flags along two dimensions: medical accuracy and patient-specific relevance. Two senior gynecologists independently evaluated a total of 335 red flags generated for 297 patients. The mean ratings were \textbf{96.30\%} for medical accuracy and \textbf{93.86\%} for patient-specific relevance.
\begin{table}[h]
\begin{tabular}{|l|r|}
\hline
Total Patients & 297 \\ \hline
Patients with one or more Red Flags & 163\\ \hline
Total Red Flags & 335 \\ \hline
Medical Accuracy & 96.30\% \\ \hline
Patient-Specific Relevance & 93.86\% \\ \hline
\end{tabular}
\centering
\caption{Overview of Red Flags Evaluation by Consultant Gynecologists}
\end{table}

\subsection{Cross-Model Evaluation}
To assess the comparative performance of System X across different LLMs, we conducted a controlled evaluation using the most recent 50 patient records. The original deployment used GPT-4o for EMR generation, with clarification questions answered by attending healthcare workers. For a fair comparison, we re-evaluated the same dataset using top LLMs: Gemini 2.5 Pro\footnote{https://deepmind.google/models/gemini/pro/}, Claude Sonnet 4.5~\footnote{https://www.anthropic.com/claude/sonnet}, DeepSeek-V3.2~\footnote{https://api-docs.deepseek.com/news/news250929}, and GPT-5~\footnote{https://openai.com/gpt-5/}, the model currently powering the deployed system. To eliminate inter-annotator variation, all clarification questions were answered by a single member of the research team using ground-truth transcriptions. We computed field-by-field accuracy across six EMR sections (Accuracy = Total correct fields/ Total fields in the EMR), following the same evaluation procedure described in Section~\ref{subsec:categorization}. As shown in Table \ref{tab:llm_comparison}, the GPT-5-based pipeline of System X remained comparable to other leading LLMs, achieving 96.4\% overall field-level accuracy. Gemini and Sonnet showed similar performance, while DeepSeek trailed slightly. Overall, the results demonstrate the reliability and modularity of System X’s EMR generation pipeline, indicating that its design grounded in domain specific prompts and clarification workflows is as critical to accuracy as the underlying model architecture or scale.

We tested several domain-specific medical LLMs~\cite{Yang2022,chen2023meditron70b,medfound} but excluded them from the comparative evaluation due to their inability to generate natural-language clarification questions from ASR-derived transcriptions. While these models are highly optimized for biomedical text understanding and concept extraction, they lack conversational prompting capabilities making them unsuitable for fair comparison against System X, where dialogue-based refinement constitutes a critical step in EMR generation.

\begin{table}[h!]
\centering
\small
\begin{tabular}{lccccc}
\toprule
\textbf{Section} & \textbf{Total Fields} & \textbf{Gemini 2.5 Pro} & \textbf{Sonnet 4.5} & \textbf{DeepSeek-V3.2} & \textbf{GPT-5} \\
\midrule
Medical History & 1568 & 96.2\% & 96.1\% & 94.2\% & \textbf{96.4\%} \\
Socio-Economic History & 270 & 96.3\% & 96.3\% & 87.8\%	& 92.2\% \\
Present Pregnancy & 850 & 95.6\% & 95.1\% & 92.8\% & 94.9\% \\
Family History & 480 & 97.7\% & 95.4\% & 92.5\% & \textbf{97.3\%} \\
Past Pregnancy & 840 & \textbf{99.5\%} & 99.0\% & 98.1\% & 98.1\% \\
Proposed Plan & 147 & 95.9\% & 96.6\% & 81.0\% & \textbf{99.3\%} \\
\midrule
\textbf{Overall Accuracy} &  & \textbf{96.9\%} & 96.4\% & 93.6\% & 96.4\% \\
\bottomrule
\end{tabular}
\caption{Cross-model comparison of EMR field-level accuracy across 50 most recent patient records}
\label{tab:llm_comparison}
\end{table}

\subsection{Qualitative Evaluation}
\label{subsec: feedbackinterviews}
\subsubsection{Methods: Participants and Procedure} As our pilot deployment neared completion and the system had collected 500+ patient records, we conducted feedback interviews with four users (P1-P4) of the application. Additionally, we interviewed a consultant gynaecologist (C1) who oversaw the deployment of the application at Hospital Y. Participants were selected based on their diverse roles and status as top users of the application. Each had consistently used the system for over a month and recorded complete patient histories for at least 20 patients, ensuring thorough familiarity of participants with the application. All interviews were conducted in person, except for one, which was held via Google Meet because of in-person unavailability of the participant. We obtained consent to record the interviews, which lasted between 15 and 45 minutes.

The interview protocol (refer to \ref{feedbackinterview}) focused on user perceptions of the usability of the application, changes in workflows before and after adopting the application, data collection practices and challenges encountered during its deployment. The interview with the consultant physician used a more open approach, focusing on the broader potential and implications of application in clinical practice.

\begin{table}[h]
\begin{tabular}{|l|l|}
\hline
\textbf{Participant} & \textbf{Designation}    \\ \hline
C1                   & Senior Consultant              \\ \hline
P1                   & Post Graduate Doctor (Senior)
       \\ \hline
P2                   & Nurse Navigator         \\ \hline
P3                   & Nurse Navigator         \\ \hline
P4                   & House Officer Doctor (Junior)      \\ \hline
\end{tabular}
\centering
\caption{Feedback Interview Participants' Designations}
\end{table}

All interviews were transcribed, and recurring patterns were identified using inductive coding. Example codes include `socioeconomic status inquiries', `increased patient interaction', and `efficiency in history-taking'. Two team members independently coded the transcripts, and the resulting themes were merged. 
\subsubsection{Findings from Qualitative Evaluation Interviews}
\label{feedback}
The findings from our interviews demonstrate improvements in patient history documentation, risk identification, and engagement among junior doctors, as well as challenges encountered during the application’s use.

\textbf{\textit{Improvement in Patient History Documentation}}
The application improved the level of detail recorded by doctors in patient history documentation. Participants reported asking more comprehensive questions, particularly about socioeconomic factors and family history. P2 stated:
\begin{quote}
    \begin{center}
    \textit{"Even after[I have stopped using] the app, I still ask about socioeconomic history and write it on the card...There is family history, for example, we usually just note diabetes as yes/no, but we started asking additional questions like which family member has it? So, we began specifying this in the recordings [and] I do this on the antenatal card now too."} (P2)
        \end{center} 
\end{quote}
Two healthcare professionals (P2, P4) specifically mentioned inquiring about domestic abuse when taking patient histories. Previously, patient histories were recorded routinely on antenatal cards, but now doctors are collecting much more detailed information, often overlooked in the past, that patients typically wouldn't disclose on their own. This approach even led to the identification of domestic violence cases. P2 explicitly recalled documenting such an instance during the interview. P3 also mentioned becoming more efficient in taking detailed patient histories, allowing her to spend more time with patients. Even after discontinuing the use of the application, she continued to collect additional information as it had become part of her routine.\\

\textbf{\textit{Improved Risk Identification}} The application's prompts facilitated the collection of more detailed patient information and. P1 noted:
\begin{quote}
    \begin{center}
    \textit{"Sometimes, details like blood group get missed, and the app highlights them. We often overlook things like cousin marriages, but the app points them out. In Pakistan, we don't focus much on genetic testing; some consultants address these issues, but not all."} (P1)
        \end{center} 
\end{quote}
P2 provided a specific example:
\begin{quote}
    \begin{center}
    \textit{"For previous pregnancies, we now document details like at what week a miscarriage or abortion occurred. A red flag prompted us to include whether the miscarriage led to a DNC or not, which is crucial information to have."} (P2)\\
        \end{center} 
\end{quote}
During a field visit, one doctor also appreciated the red flag feature for highlighting the importance of maternal vaccination in preventing conditions such as pneumonia in newborns. This red flag was particularly important for a patient who had previously lost a child to the illness.\\

\textbf{\textit{Higher Engagement Among Junior Doctors}} Based on the collected data and feedback interviews, we observed that the acceptability of the application varied by the doctors' seniority levels. Trainee doctors, with fewer external duties, were more available to record complete data and respond to transcription clarifications and additional medical questions. In contrast, senior doctors, who had responsibilities overseeing the labor ward and assisting in surgeries, often left medical questions unanswered. During the interviews, trainee doctors showed greater motivation to continue using the application. 

Junior doctors and trainees liked features such as red flag generation and additional medical questions. For example, junior doctors [P2, P3] found it helpful when the system flagged previous medical events like miscarriages or congenital issues, prompting a thorough review of patient history and potential follow-up treatments. In contrast, senior doctors, with their extensive experience, found less novelty in these features, indicating that the application's impact may vary based on the user's level of experience.\\

\textbf{\textit{Challenges with the Application’s Use}} The most commonly reported challenges across interviews were transcription errors and the effect of background noise on accuracy. These issues forced doctors to often correct transcriptions or edit EMRs, increasing the time required to capture patient data. Internet connectivity was also a consistent issue during deployment. The hospital did not provide internet access, and the devices we supplied performed poorly in the outpatient department. This caused frequent server connection failures and delays in generating transcriptions and EMRs, leading to user frustration from repeated error notifications.

\section{Discussion}
Our application, piloted and deployed at Hospital Y for approximately seven months, successfully generated over 500 EMRs for new patients and 42 EMRs for returning pregnant patients. In this section, we discuss key insights from the pilot and to offer potential frameworks and pathways to creating AI based health technologies for low resource countries.

\subsection{Bridging Linguistic Gaps: The Necessity of Contextual Medical Dictionaries for Accurate LLM Deployment in Healthcare}

In our first iteration of using the LLM our knowledge base consisted only of the medical guidelines \cite{rcog}. However, during analysis of the first few generated EMRs it became clear that there it is critical to curate a customized dictionary from existing patient records and interview data, that can bridge the gap between colloquial medical terms commonly used in Pakistan and their corresponding medical terminologies. This is an important consideration for systems like ours where healthcare workers frequently use informal, indigenous terms to describe medical conditions. For example, the term `sugar' is often used to refer to `diabetes' in Pakistan. While this is widely understood in a local context, our LLM does not understand this nuance which leads to incorrect EMRs. Similarly, terms such as `blood pressure' may be used to refer to hypertension. This dictionary mapping is particularly important in settings where healthcare workers may lack formal training in medical terminology and rely on colloquial expressions. This dictionary will vary across different contexts and languages.

\subsection{Designing for Accurate Medical Records through Speech}


In order to ensure complete and accurate EMRs, we introduced a brief dialogue with the LLM that allows health care workers to answer follow up questions that the LLM might have about incomplete or incorrect or confusing data inputs.

A critical challenge in generating accurate EMRs from speech-based transcriptions in a hospital setting is the potential for errors or omissions in the transcription process. In noisy, high-pressure environments such as hospitals, it is common for words to be incorrectly transcribed, misspelled, or for important details to be missed altogether. These errors can significantly impact the quality of the EMR and, by extension, the clinical decisions made based on that data.

In our system the LLM detects ambiguities, missing information, or potential errors in the transcription, it prompts the healthcare professional with targeted clarification questions. These questions serve to confirm or correct key details, ensuring the EMR is complete and accurate. For instance, if the system identifies an outlier in a patient’s vital signs or a potential omission in medication records, it may ask, \textit{`Can you confirm the patient’s current blood pressure?'} or \textit{`Is the patient still prescribed anti-hypertensive medication?'} This process helps to safeguard against errors that could otherwise be overlooked in noisy environments or during busy clinical workflows.

 Clarification questions also help reduce the cognitive burden on healthcare professionals by focusing their attention on areas of uncertainty rather than requiring full transcription reviews. Given that many healthcare workers in low-resource settings operate with limited time and may not have extensive training in medical record-keeping, prompting them with specific, actionable questions minimizes the mental effort required while improving the quality of the recorded data.

\subsection{Designing LLMs for Diagnostic Support}

\subsubsection{Improving Diagnostic Outputs: Leveraging Retrieval-Augmented Generation (RAG) for Improved Decision Support in Low-Resource Settings}

In our initial implementation, the LLM successfully generated red flags by identifying key conditions such as "BMI high" or "hypertension", which were medically correct and aligned with the information captured in the EMRs. These red flags though factually correct tended to be long, giving detailed explanations of terms like BMI and providing generic information which during evaluation, doctors did not have the time or desire to read. As an ongoing work and to further improve the system, we integrated a Retrieval-Augmented Generation (RAG) framework into the system. The RAG approach allowed the LLM to retrieve relevant maternal healthcare guidelines from trusted sources such as the National Institutes of Health (NIH) and the Royal College of Obstetricians and Gynecologists (RCOG) during the generation of these diagnostics which were more specific, useful and better structured.

The addition of the RAG framework enhanced the system's ability to provide more context-specific and detailed diagnostics. Instead of just flagging conditions, the system now offers richer clinical insights, including recommendations for further testing, monitoring protocols, and medication reviews. These enhanced outputs are particularly valuable in low-resource settings where healthcare workers may benefit from more comprehensive decision support, helping them make informed choices even in the absence of extensive training. Table \ref{detailsRag} shows an example of red flags raised before and after RAG 
\begin{table}[h]
\centering
\renewcommand{\arraystretch}{1.5} 
\begin{tabular}{|>{\hspace{0.3cm}}p{6cm}<{\hspace{0.3cm}}|>{\hspace{0.3cm}}p{7cm}<{\hspace{0.3cm}}|}
\hline
\textbf{Prompted GPT-4o} & \textbf{Prompted GPT-4o with RAG} \\ \hline

\textbf{1. BMI Above 30 kg/m²:}  
The patient's BMI is 37.9, indicating obesity, which poses several risks in pregnancy, including gestational diabetes, hypertension, and increased likelihood of C-section delivery. 

\textbf{2. Hypertension:}  
The patient has a medical history of hypertension and is currently on Norvasic, pointing to chronic hypertension, which must be closely managed to reduce risks such as preeclampsia, intrauterine growth restriction (IUGR), and preterm birth.

\textbf{3. Family History:}  
Positive family history of diabetes and hypertension on the patient’s side, and multiple pregnancies on the husband’s side, which is important for risk stratification.

\textbf{4. Present Pregnancy:}  
No reported symptoms of burning micturition, PV discharge, itching, nausea, vomiting, bowel issues, contractions, or vaginal bleeding. Food intake is good, and the abdomen is soft and non-tender. 
&
\textbf{1. Hypertension:}  
The patient has a history of hypertension, requiring regular monitoring. Blood pressure should be checked regularly, ensuring readings are within safe limits (135/85 mmHg or less) if on antihypertensive treatment.

\textbf{2. Gestational Diabetes Monitoring:}  
Given the family history of diabetes and the patient's obesity (BMI of 37.9), screening for gestational diabetes is crucial. Conduct an HbA1C test and a random blood glucose test.

\textbf{3. Obesity-Related Risks:}  
The patient's high BMI increases risks such as preeclampsia and gestational diabetes. A detailed nutritional and physical activity plan should be developed.

\textbf{4. Fetal Monitoring:}  
Regular ultrasound scans are necessary to monitor fetal growth, especially given the history of a large baby (4 kg at 37 weeks). Fetal movements should be monitored.

\textbf{5. Medication Review:}  
The patient is currently on Norvasc and Myabolic Acid. Review these medications for safety during pregnancy, and consult an obstetrician for adjustments if needed.

\textbf{6. Cardiac Assessment:}  
Given the patient's hypertensive status and family history, a cardiologist referral may be needed to rule out any underlying cardiac conditions. 
\\ \hline

\end{tabular}
\caption{Comparison of Diagnostics Generated Before and After RAG Integration. The system prompt was kept identical across both configurations to ensure a consistent basis for comparison of diagnostic outputs.}
\label{detailsRag}
\end{table}

While the initial system delivered medically accurate diagnostics in the setting where it was first deployed, the enhanced model further strengthens its utility by providing more detailed, actionable insights, making it even more effective in diverse and low-resource healthcare environments.

\subsubsection{Impact of RAG on Diagnostic Quality}

The RAG framework improved diagnostic outputs in several key ways: \begin{itemize} \item \textbf{Increased Diagnostic Depth:} The enhanced diagnostics provided not just a simple flagging of issues like high BMI or hypertension, but reasoned insights that drew on clinical guidelines. This allowed the system to suggest specific interventions, such as HbA1C testing or cardiac referrals, based on the patient’s overall condition and risk factors. \item \textbf{Actionable Insights for Healthcare Workers:} In resource-limited settings, where healthcare workers may have minimal training, the ability to generate contextually aware and actionable insights is critical. RAG-enhanced outputs helped guide clinical decision-making by providing steps for further investigation and management. \item \textbf{Moving Beyond EMR Recap:} Unlike the initial approach, which largely echoed information already contained in the EMRs, the RAG-enhanced diagnostics synthesized data and applied medical guidelines to generate more informed, reasoned conclusions. This not only added value to the diagnostic process but also ensured that important clinical considerations were not overlooked. \item \textbf{Contextualized Risk Assessment:} By referencing authoritative guidelines, the system provided a more nuanced assessment of patient risk. For instance, rather than merely flagging a high BMI, the enhanced output recommended managing weight gain and conducting additional monitoring, helping healthcare workers mitigate risks in real-time. \end{itemize}

The integration of the RAG framework allowed for a significant enhancement in the quality of diagnostics generated by our system. This improvement is particularly important in low-resource settings, where the system's ability to offer actionable, clinically informed diagnostics can directly impact patient outcomes by empowering healthcare workers with better decision support.

\subsubsection{Leveraging Structured Outputs for Consistent and Accurate EMRs}

In our initial approach, simply prompting the LLM with a template for generating EMRs proved to be insufficient. The LLM would sometimes deviate from the prescribed format, resulting in issues such as missing fields, incorrect data placement, and occasional misspellings. These inconsistencies created significant challenges, particularly when it came to analyzing the EMRs or extracting key information. Due to the variability in structure, using regular expressions (regex) to extract specific fields became increasingly unreliable.

To resolve this issue, we integrated the recently introduced structured output feature of GPT-4, allowing us to define a JSON schema for each EMR template. By adhering to this well-defined schema, the system produced consistently formatted EMRs with structured key-value pairs. The introduction of structured outputs had several key benefits:

\begin{itemize}
    \item \textbf{Increased Accuracy}: By enforcing schema adherence, the LLM generated accurate and complete EMRs, significantly reducing errors and inconsistencies.
    \item \textbf{Ease of Analysis}: The structured nature of JSON objects made it much easier to query and analyze EMRs compared to free-text outputs. Unlike textual formats, which require complex pattern matching, JSON allows straightforward data extraction and manipulation.
    \item \textbf{Reducing the Risk of Data Loss}: Inconsistent or inaccurate EMRs can result in the omission of critical patient information. This poses a particular risk in maternal healthcare, where accurate data is crucial for identifying potential complications. Structured outputs ensure that essential fields, such as symptoms, diagnoses, medications, and social determinants of health, are consistently captured.
\end{itemize}

By leveraging structured outputs, we significantly improved the quality and reliability of the generated EMRs, ensuring that healthcare workers have access to accurate and complete patient data. This structured approach also facilitates more efficient analysis, which is critical for making timely and informed decisions in maternal healthcare settings.

\subsection{Framework for Evaluating LLM-Based Applications in Healthcare}
There is currently a lack of a comprehensive framework for evaluating LLM-based applications, particularly in the healthcare context. Traditional performance metrics, while useful, primarily assess the capabilities of the LLM itself, rather than the overall effectiveness of the application. For instance, ChatGPT’s performance on the United States Medical Licensing Exam (USMLE), where it scored at or near the passing threshold, demonstrates potential for supporting medical education and clinical decision-making \cite{kung2023performance}. However, relying on such metrics alone can lead to misleading conclusions about a system’s real-world impact and does not account for the nuances of using such a system in a low-resource setting with unskilled health care workers.  

In high-stakes domains such as healthcare, achieving high accuracy is essential, but assessing the potential impact of system errors is equally important. Relying solely on accuracy as a success metric is insufficient. To evaluate the correctness of system-generated EMRs, we not only measured accuracy but also analyzed the potential for patient harm associated with each error. We developed three categories to reflect how healthcare workers would address inaccuracies: they might either ignore insignificant errors, easily identify and correct them, or verify the actual information by reviewing the patient’s record. Errors categorized as "No Action Needed" or "Easily Identifiable and Correctable" do not result in harm to the patient. We extended this dual evaluation approach to the red flag feature, assessing both its medical accuracy and its utility for doctors. Building on our deployment experience, we propose a framework for developing context-specific evaluation metrics. Table \ref{tab:evaluation-framework} outlines evaluation steps, what data is required to carry out evaluation and what is the output of that evaluation. For evaluations requiring domain expertise, we recommend involving two evaluators to enhance the validity and reliability of the assessment.
\begin{table}[htbp]
\resizebox{\textwidth}{!}{%
\begin{tabular}{|p{3.5cm}|p{4cm}|p{5cm}|p{5cm}|}
\hline
\textbf{Evaluation Step} & \textbf{Input: What is required for evaluation at this step?} & \textbf{Output: What specific results or findings does this step generate?} & \textbf{Purpose: What is the need of this step?} \\ \hline

\textbf{Effectiveness of speech-to-text Model} \newline (Evaluated by research team) & 
(1) Manually transcribed audio recordings to generate "ground truth" transcripts \newline (2) System-generated transcripts from ASR model & 
(1) WER percentage: Calculates errors in transcriptions by comparing with ground truth\newline(2) Dataset to build a medical dictionary of mis-transcribed terms. & 
(1) To assess transcription quality of the speech-to-text model for patient antenatal medical data \newline(2) Improve understanding of indigenous terms in healthcare.\\ \hline

\textbf{Correctness of system-generated EMRs}\newline (Evaluated by research team) & 
(1) Ground-truth EMRs from manual transcription \newline
(2) System-generated EMRs & 
(1) Number of correctly filled template fields.\newline
(2) Overall accuracy of EMR generation. & 
(1) To evaluate the correctness of generated EMRs and analyze errors \newline
(2) Identify frequently misfilled fields and adjust system prompts. \\ \hline

\textbf{Categorization of EMR Inaccuracies}\newline (Evaluated by Senior Consultants) & 
(1) Incorrect fields in system-generated EMRs\newline 
(2) Ground-truth EMRs\newline
(3) Expert review by gynecologists & 
Categorization into three categories: \newline(1) No Action Needed\newline (2) Easily Identifiable and Correctable\newline (3) Uncorrectable Without Ground Truth & 
(1) Understand clinical significance of inaccuracies and potential harm that could reach the patients\newline
(2) Understand healthcare workers' response to inaccuracies \\ \hline

\textbf{Red Flag Evaluation} \newline(Evaluated by Senior Consultants) & 
(1) System-generated red flags for each patient\newline 
(2) Expert review by gynecologists & 
(1) Mark as 'Medically accurate' or 'Medically inaccurate' \newline
(2) Assess patient-specific relevance \newline
(3) Identify missing red flags & 
(1) Evaluate accuracy and relevance of red flags\newline
(2) Add guidelines to improve patient-specific red flags \\ \hline
\end{tabular}
}
\caption{Evaluation Framework for Speech-to-EMR System}
\label{tab:evaluation-framework}
\end{table}

\section{Future Work}
Currently, the system aids doctors and nurse practitioners in creating EMRs via speech and provides diagnostic support by flagging potential issues. This can be particularly beneficial for midwives and community health workers in peri-urban clinics, where standardized patient record systems are often lacking. Our next phase aims to extend the system to maternity homes and similar settings. While currently focused on antenatal care and EMR management for doctors, future iterations could include patient-facing features, allowing individuals to access their own records and care plans. This could enhance adherence to care protocols and improve maternal health outcomes in under-served areas.

\section{Conclusion}
Our pilot deployment establishes a foundation for implementing AI-powered healthcare systems in resource-constrained settings such as Pakistan. By incorporating speech-based input in the local language, the system demonstrated the potential to enhance maternal healthcare through efficient digital record-keeping. Generating over 500 EMRs, the application’s design effectively bridged the gap between low digital literacy and advanced healthcare needs, supporting doctors and nurses with intelligent prompts to facilitate comprehensive patient data collection. Notably, the system captured patients' socioeconomic history, a critical but previously overlooked component of patient records, as identified by doctors and addressed a major risk associated with paper-based antenatal cards, which often led to data loss when misplaced. A key component of our system was the generation of `red flags' based on patient data to assist doctors in identifying potential concerns early during the patient's pregnancy.

In the absence of established evaluation protocols, we developed a comprehensive framework involving both the technical research team and domain experts. However, challenges remain, particularly the limited availability of speech data in regional languages and usability issues in noisy environments. These obstacles highlight the need for further iterations and system improvements. Future work should focus on enhancing the accuracy of the ASR and LLM modules and exploring patient-facing functionalities to improve engagement and adherence to care plans. The insights from this study provide valuable guidelines for deploying and evaluating AI-driven solutions in similar low-resource settings, emphasizing the importance of context-aware design and the integration of local healthcare practices to achieve meaningful impact.

\bibliographystyle{ACM-Reference-Format}
\bibliography{maternalhealth-base}

\appendix
\section{Appendices}

\subsection{Example of Medical Questions}\label{subsec:medquestions}
1. What is the husband's blood group? \\
2. What is the reason for the prescription of Iron and Calcium tablet? Does the patient have any past issues related to Iron and Calcium deficiency? \\
3. As the patient is in a consanguineous marriage, are there any inherited diseases in their families that should be monitored? \\
4. How long has the patient had no vaginal bleeding since her last period, considering she is currently pregnant? \\
5. Does the patient follow any specific diet or pursue physical activities? \\
6. Are there any aspects of the patient's work as a Staff Nurse that could affect her pregnancy, such as her working hours, stress levels, or exposure to certain environments or substances? \\
7. Does the patient have any history of mental health issues, such as anxiety or depression? \\

\subsection{Medical Dictionary}\label{subsec:med-vocab}
\begin{multicols}{2}
\begin{enumerate}
    \item Gravida
    \item Para
    \item Abortions
    \item Cytology
    \item Consanguineous Marriage
    \item Puerperium
    \item Edema
    \item Weeks of Gestation
    \item Burning Micturition
    \item PV Discharge
    \item EDD (Expected Delivery Date)
    \item LMP (Last Menstrual Period)
    \item Sugar (referring to Diabetes)
    \item Blood Pressure (referring to Hypertension)
    \item Dil ka marz (referring to Heart Disease)
    \item Smoking
    \item Nasha (referring to Substance Abuse)
    \item Fundal Height
    \item Fetal Parts Palpable
    \item Engagement
    \item Fetal Movements
    \item Nausea
    \item Vomiting
    \item Diarrhea
    \item Constipation
    \item Contractions
    \item Vaginal Leaking or Discharge
    \item Water Bag Leakage
    \item Vaginal Bleeding
    \item Spontaneous Vaginal Delivery
    \item Complication
    \item Iron Calcium Therapy
    \item Base Lines
    \item CBC (Complete Blood Count)
    \item Urine Complete Examination
    \item Hepatitis B and C
    \item BSR (Blood Sugar Random)
    \item OB Scan (Obstetric Scan)
    \item Antituberculous Therapy
    \item Blood Transfusion
    \item Tetanus Toxoid
    \item Cardiac Problem
    \item Cephalic Presentation
    \item Cephalic
    \item Investigations
    \item Serum Ferritin
    \item HB Electrophoresis
    \item Hb (Hemoglobin)
    \item Iron
    \item Calcium
    \item Tablet Loprin
    \item Abdominal Pain
    \item Pehla hamal (referring to First Pregnancy)
    \item UPT (Urine Pregnancy Test)
    \item Cesarean Section
    \item Failed Induction of Labor
    \item Spontaneously Conceive
    \item Folic Acid
    \item Oral Contraceptives
    \item SVD (Spontaneous Vaginal Delivery)
    \item Duration of Labor
    \item Maturity in Weeks
    \item Cran Max
    \item Postnatal Complication
    \item Fetal Anomaly
    \item DNC (Dilation and Curettage)
    \item DVT (Deep Vein Thrombosis)
    \item IGF (Intrauterine Growth Restriction)
    \item Weight of the Baby
    \item Alive and Healthy
    \item Mode of Delivery
    \item Height (referring to Fundal Height)
    \item Longitudinal Cephalic
    \item Risk Factors
    \item Pre-eclampsia
    \item Eclampsia
    \item PIH (Pregnancy Induced Hypertension)
    \item Recurrent Miscarriages
    \item Endocrine Dysfunction
    \item Infertility
    \item Anesthetic Problems
    \item Multiple Pregnancies
    \item Hemoglobinopathy
    \item Congenital Anomaly
    \item TB (Tuberculosis)
    \item HbA1c (Glycated Hemoglobin)
    \item Abdominal Wall Thickness
    \item Placenta Previa
    \item Term
    \item Postnatal Complications
    \item Puerperium Complications
    \item Depression (referring to Postpartum Depression)
    \item Sepsis
    \item RH Incompatibility
    \item Symphysio Fundal Height
    \item Itching in the Vaginal Area
    \item Leakage of Fluid per Vagina
    \item Sono Engagement
    \item Oligohydramnios
    \item Breech Presentation
    \item Cerclage
    \item Anemia
    \item Postpartum Depression
    \item Dexa Scan
    \item Dexa Cortisone (referring to Dexamethasone)
    \item Tranexamic Acid Injections
    \item Vitamin D
    \item Postpartum Hemorrhage
\end{enumerate}
\end{multicols}

\subsection{Initial Semi-Structured Interview Protocol}
\label{initialprotocol}
\begin{enumerate}
    \item How do you decide which information to share with the patient, and when?
    \item Are there any differences between how you approach this in public versus private practice?
    \item How do you assess a patient’s educational status during a consultation?
    \item What indicators do you use to determine their understanding of the treatment or care plan?
    \item Can you describe the process for recording vital signs such as weight, BMI, and blood pressure during pregnancy?
    \item What tools or charts do you use to monitor fetal growth, and do you think they are suitable for the local population?
    \item What lab tests are commonly recorded, and how is this information stored?
    \item Are there any tests that are outdated or should be removed from the standard antenatal record?
    \item How do you schedule follow-up visits with your patients? Do you use a calendar system?
    \item How consistent are patients in attending their follow-up visits?
    \item At what point in the patient’s journey is it confirmed that she is pregnant?
    \item What diagnostic methods do you use if the patient does not know she is pregnant?
    \item How frequently do patients miss scheduled visits, and what steps are taken to retrieve their medical records when they return?
    \item Can you describe the order in which you complete an antenatal card during a consultation?
    \item Are there any parts of the card that you find unnecessary or challenging to complete?
    \item Do you ask patients about socioeconomic factors? If so, when during the consultation?
    \item What additional clinical symptoms (e.g., burning micturition, vaginal discharge) do you think should be included or excluded from the antenatal record?
\end{enumerate}

\subsection{Antenatal Card - Front Page}\label{subsec: antenatalcard} 
\begin{figure}[H]
\centering
  \includegraphics[width=0.65\textwidth]{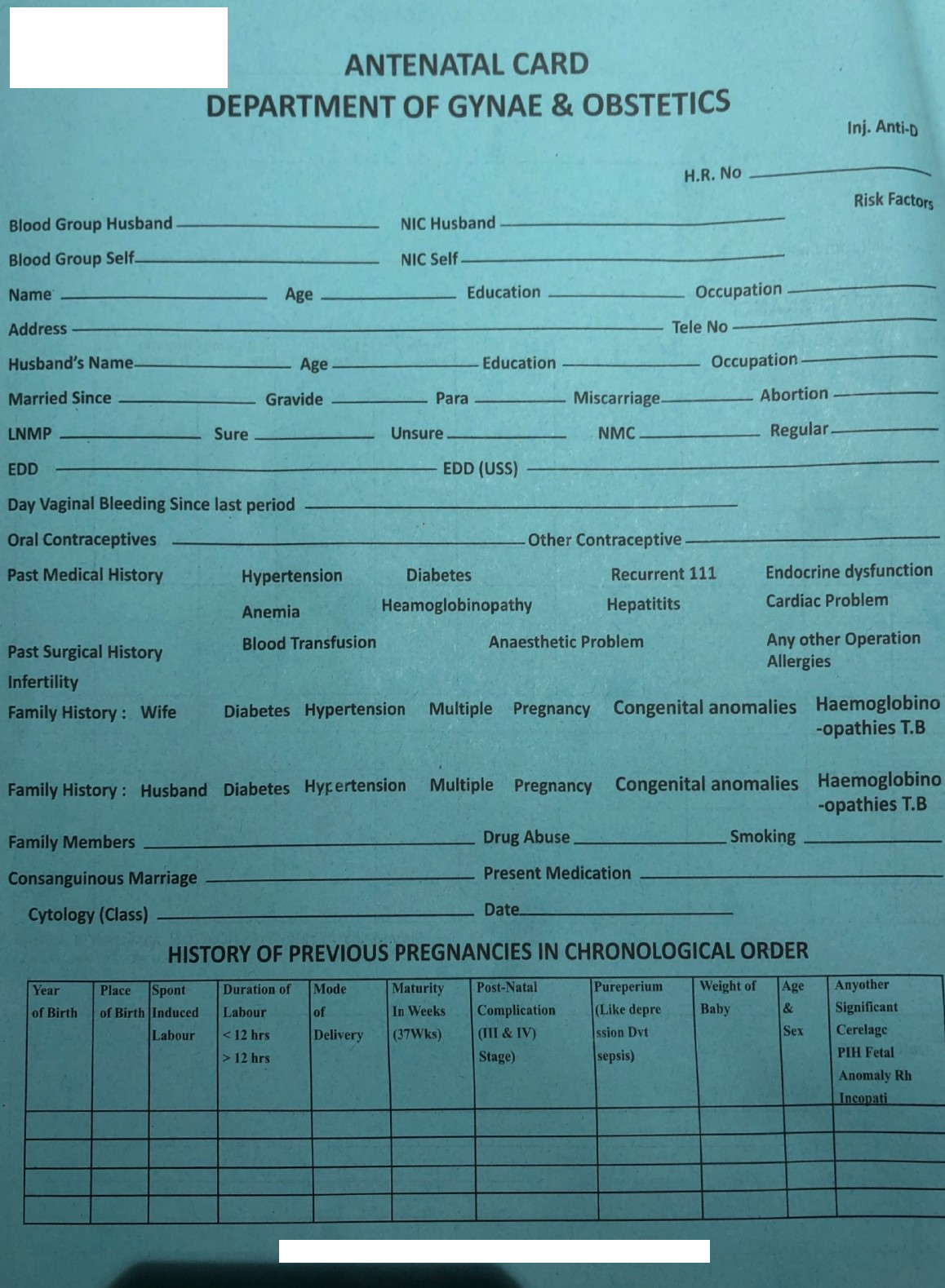}
  \caption{Front page of the antenatal card used at Hospital Y}
\end{figure}

\subsection{Qualitative Evaluation Interview Protocol}
\label{feedbackinterview}
\begin{enumerate}
    \item How long have you been using the application?
    \item What changes have you observed from September/October when you started data collection to the latest February/March application version?
    \item Did you find the app easier to use as it kept changing? 
    \begin{itemize}
        \item If yes, in what ways?
        \item If not, in what ways?
    \end{itemize}
    \item Have you noticed any improvements in the data being collected from the patient before vs after using the application? (Assess if the additional information asked from a patient in the app’s EMR is incorporated in their current workflows)
    \item Did asking for additional information like socioeconomic status and concerns/risk factors contribute towards better patient outcomes? (such as having more complete information or having this information led to better decisions?)
    \item How has your workflow changed since incorporating the app into your practice?
    \item What were the biggest challenges when using the application?
    \item Can you recall any particular instance where you found the additional questions useful? Vice versa.
    \item Can you recall any particular instance where you found the red flag raised useful and something that you had not thought of yourself? Vice versa.
    \item What was your overall experience using the application? (Keep this open-ended to assess if they had a positive or negative experience; probe further based on the reply.)
    \item Based on your experience using the application, do you have any recommendations for its improvement?
    \item Is there anything else you want to share about your experience using the app, or do you have any additional comments?
\end{enumerate}

\subsection{The application's screen for recording present pregnancy details}\label{subsec:app-screen}

\begin{figure}[h]
\centering
  \includegraphics[width=0.5\linewidth]{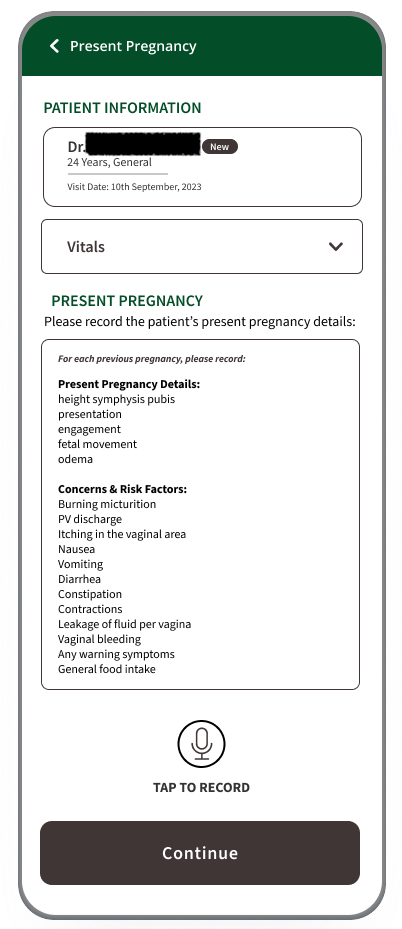}
  \caption{The application's screen for recording present pregnancy details}
\end{figure}

\subsection{Example Generated EMR of a Patient}\label{subsec:emr-template}

\begin{table}[h]
  \centering
  \begin{tabular}{|p{5cm}|p{7cm}|}
    \hline
    \textbf{CLINICAL PARAMETER} & \textbf{DATA} \\
    \hline
    \textbf{Personal Details} & \textbf{} \\
    Blood Group Self & A positive \\
    Education Self & BA \\
    Occupation Self & Housewife \\
    Married Since & 1 and a half years \\
    Gravida & Primary \\
    Para & Not provided \\
    Miscarriage & No \\
    Abortion & No \\
    Normal Menstrual Cycle (Regular/Irregular) & Regular \\
    Consanguinous Marriage & No \\
    Vaginal Bleeding Since Last Period & Yes \\
    Oral Contraceptives, Other Contraceptives & No \\
    Drug Abuse & No \\
    Smoking & No \\
    Present Medications & Iron, Calcium, Loprin \\
    \hline
    \textbf{Husband Details} & \textbf{} \\
    Education & Matric \\
    Blood Group & O Positive \\
    Occupation & Businessman \\
    \hline
    \textbf{Medical History} & \textbf{} \\
    Diabetes & No \\
    Recurrent UTI & No \\
    Cardiac Problem & No \\
    Hepatitis & No \\
    Anemia & No \\
    Hypertension & No \\
    Hemoglobinopathy & No \\
    Endocrine dysfunction (Thyroid/PCOS) & No \\
    \hline
    \textbf{Surgical History} & \textbf{} \\
    Blood Transfusion & No \\
    Anesthetic Problem & No \\
    Infertility & No \\
    Operation Allergies & No \\
    \hline
    Additional Information & Patient X is 23 years old. She has been married out of the family for 1 and a half years. No history of miscarriage, abortion, oral contraceptives, other contraceptives, drug abuse, or smoking. Present medication includes iron, calcium, and loprin. Her husband, Y, is 29 years old, has a matric level education, blood group O positive, and is a businessman. No medical history of hypertension, diabetes, recurrent UTI, anemia, hepatitis, cardiac problems, hemoglobinopathy, or endocrine dysfunction. No past surgical history or history of blood transfusion, anesthesia problem, infertility, or operational allergies. \\
    \hline
  \end{tabular}
  \caption{Medical History EMR}
  \label{tab:mytable}
\end{table}

\begin{table}[h]
  \centering
  \begin{tabular}{|p{5cm}|p{7cm}|}
    \hline
    \textbf{CLINICAL PARAMETER} & \textbf{DATA} \\
    \hline
    \textbf{Patient Family History} & \textbf{} \\
    Diabetes & No \\
    Hypertension & Yes \\
    Multiple Pregnancy & No \\
    Haemoglobinopathies T.B & No \\
    Congenital Anomalies & No \\
    \hline
    \textbf{Husband Family History} & \textbf{} \\
    Diabetes & Yes \\
    Hypertension & Yes \\
    Multiple Pregnancy & No \\
    Haemoglobinopathies T.B & No \\
    Congenital Anomalies & No \\
    \hline
    Additional Information & Patient and her husband do not have a history of multiple pregnancies, no history of hemoglobinopathy, or any congenital anomalies. In the patient's family, there is no history of diabetes, but hypertension is present. In contrast, the husband's family has a history of both diabetes and hypertension. Both do not have a history of TB. \\
    \hline
  \end{tabular}
  \caption{Family History EMR}
  \label{tab:clinicaltable}
\end{table}

\begin{table}[h]
  \centering
  \begin{tabular}{|p{5cm}|p{7cm}|}
    \hline
    \textbf{CLINICAL PARAMETER} & \textbf{DATA} \\
    \hline
    No of Family Members & 20 \\
    Joint/Nuclear Family & Joint \\
    Husband Status & Living with patient \\
    Relationship with Family and Husband & Good \\
    History of Domestic Abuse & No history \\
    Mental Health Issues & No \\
    Additional Information & No additional observation \\
    \hline
  \end{tabular}
  \caption{Socioeconomic History EMR}
  \label{tab:sociotable}
\end{table}

\begin{table}[ht]
  \centering
  \begin{tabular}{|p{5cm}|p{7cm}|}
    \hline
    \textbf{CLINICAL PARAMETER} & \textbf{DATA} \\
    \hline
    \textbf{Present Pregnancy} & \textbf{} \\
    Height Symphysis Pubis & 36 cm \\
    Presentation & Longitudinal Cephalic \\
    Engagement & Head Not Engaged \\
    Fetal Movement & Present \\
    Edema & None \\
    \hline
    \textbf{Concerns and Risk Factors} & \textbf{} \\
    Burning Micturition & No \\
    PV Discharge & No \\
    Itching in Vaginal Area & No \\
    Nausea & No \\
    Vomiting & No \\
    Diarrhea & No \\
    Constipation & No \\
    Contractions & No \\
    Leakage of Fluid per Vagina & No \\
    Vaginal Bleeding & No \\
    Warning Symptoms & No \\
    General Food Intake & Satisfactory \\
    \hline
    Additional Information & The patient is currently not experiencing any reported symptoms and has satisfactory general food intake. \\
    \hline
  \end{tabular}
  \caption{Present Pregnancy EMR}
  \label{tab:presenttable}
\end{table}

\begin{table}[h]
  \centering
  \begin{tabular}{|p{5cm}|p{7cm}|}
    \hline
    \textbf{CLINICAL PARAMETER} & \textbf{DATA} \\
    \hline
    Proposed Plan & The patient is advised to control her diet and record her blood sugar levels at home. She is also suggested to go for a follow-up after one week without any referral to any other doctor. \\
    \hline
    Differential Diagnosis & No Info \\
    \hline
    Prescriptions & The patient is recommended to continue her intake of calcium and iron. \\
    \hline
    Referrals to Other Doctors & No \\
    \hline
    Additional Information & The patient is a primigravida woman at 35+3 weeks of gestation. Her fundal height is 36 cm, and the obstetric scan shows that the liquor is 17 cm, which is on the upper limit. Hence she is also advised to undertake an HbA1c test. \\
    \hline
  \end{tabular}
  \caption{Proposed Plan EMR}
  \label{tab:plan-table}
\end{table}
\end{document}